\documentclass[a4paper,11pt]{article}
\pdfoutput=1 

\usepackage{jcappub} 

\usepackage{appendix}

\usepackage[T1]{fontenc} 
\usepackage{graphicx} 
\usepackage{subcaption}
\usepackage{graphicx}

\title{\boldmath Binary-boosted Dark Matter}

\author[a,b]{Javier F. Acevedo,}
\author[a]{Adam Ritz}

\affiliation[a]{Department of Physics and Astronomy, University of Victoria, Victoria, BC V8P 5C2, Canada}
\affiliation[b]{TRIUMF, Vancouver, BC V6T 2A3, Canada}

\emailAdd{jfacev@uvic.ca}
\emailAdd{aritz@uvic.ca}

\abstract{We explore the aggregate effect of binary systems on the Milky Way's dark matter (DM) velocity distribution with Monte Carlo simulations. Through gravitational interactions with binaries, transiting DM particles can gain substantial energy. We analyze this mechanism across a range of galactic binaries, and find it to be most effective for double black holes, where ejection speeds can reach  $\sim 2000 \ \rm km/s$ while attaining a large ejection rate. We assess the expected binary-boosted DM flux from synthetic populations of black hole binaries in the galaxy, and show direct detection experiments can be sensitive to it. In particular, we demonstrate that large noble liquid detectors such as Lux-Zeplin and PandaX-4T can extend their mass sensitivity down to the sub-GeV scale, and potentially become competitive with other lower-threshold experiments when the full galactic black hole binary population is taken into account. This boosting mechanism, being gravitational in nature, is largely model- and mass-independent.}

\begin{document}
\maketitle
\flushbottom

\section{Introduction}
Dark matter (DM) direct detection experiments searching for nuclear recoil constitute a powerful probe of particle DM models at the GeV scale and above. In this regime, noble-liquid detectors have achieved unprecedented sensitivity, now at the level of observing elastic coherent neutrino-nucleus interactions of solar neutrinos \cite{LZ:2025igz}. However, their sensitivity to DM-nucleus scattering rapidly degrades for lighter DM candidates, as the combination of a lower mass with the finite velocity of particles in the halo implies recoil energies that fall below observational thresholds. To overcome this limitation, a variety of alternative detection strategies have been developed to access sub-GeV DM masses, see $e.g.$ \cite{Kahn:2021ttr}. While these can be successful in extending experimental reach, such approaches are generally constrained by smaller detector volumes, leading to reduced exposure and therefore sensitivity. 

A distinct possibility is that the nature of DM itself enhances detectability prospects in the sub-GeV regime, by boosting some fraction of the halo particles to higher speeds. Such scenarios have been extensively explored in the literature, including models in which DM is accelerated through up-scattering by cosmic rays \cite{Bringmann:2018cvk,Ema:2018bih,Alvey:2019zaa,Cappiello:2018hsu,Cappiello:2019qsw,Bell:2021xff,Maity:2022exk,Bell:2023sdq}, neutrinos \cite{Zhang:2020nis,Das:2021lcr,Lin:2022dbl} and supernovae \cite{Cappiello:2022exa}, interactions in the solar interior \cite{Kouvaris:2015nsa,Emken:2017hnp,An:2017ojc,Emken:2021lgc,An:2021qdl,Emken:2024nox}, emission from blazars \cite{Granelli:2022ysi}, or simply from the dynamics of a complex dark sector \cite{Berger:2014sqa,Agashe:2014yua,Giudice:2017zke,Geller:2022gey,Acevedo:2024wmx}. While these mechanisms can significantly enhance detector sensitivity, they often rely on specific assumptions about the interaction of DM with the Standard Model (SM), require nontrivial dark sector dynamics, or suffer from a strong suppression in the number density of the boosted component. In many cases, the resulting signal is model-dependent and comes at the expense of a reduced flux.

In this work, we introduce a qualitatively different boosting mechanism. We demonstrate that gravitational interactions between DM particles and binary systems of black holes and stars in the Milky Way can impart substantial kinetic energy to particle DM, continuously generating a population of fast-moving particles with velocities well above the conventional halo velocity cutoff. As this process only relies on gravity, it is therefore largely independent of the DM particle mass and of its non-gravitational interactions, thus avoiding many of the shortcomings inherent to the aforementioned scenarios. Based on simple analytic arguments, we predict that short-period double black hole binaries are especially effective for this mechanism, both in terms of the up-scattering rate and the energy reach. We subsequently confirm this expectation utilizing Monte Carlo simulations constructed entirely from first principles. 

As a direct application of this process, we estimate the boosted DM distributions arising from a few sample populations of black hole binaries in the Milky Way. These are computed using synthetic catalogues that incorporate the Galaxy’s star formation history. We also identify the Milky Way's nuclear star cluster as potentially the most dominant source of binary-boosted DM, both in terms of flux and energy of the ejecta. Using these distributions, we then assess observability in direct detection experiments. Remarkably, we find that large-volume noble-liquid detectors such as Lux-Zeplin (LZ) \cite{LZ:2015kxe} and PandaX-4T \cite{PandaX:2018wtu} can be sensitive to this gravitationally-boosted DM component through nuclear scattering. We estimate the extension to their mass reach for benchmark spin-independent and spin-dependent interactions, and find that it can reach down to $\sim 500 \ \rm MeV$. In addition, we show how the sensitivity to inelastic DM scattering can be also extended even for DM at the TeV scale, a unique feature of this boosting channel owing to its mass-independence. While there are astrophysical uncertainties concerning the abundance and properties of black hole binaries in the Milky Way, these results illustrate how this new process could enable this existing class of detectors to robustly probe masses well below their reported limits. 

The structure of this paper is as follows: we first review a simplified analytic description for the energy gain through gravitational scattering in Section~\ref{sec:grav_ejection_analytic}. Section~\ref{sec:sim_description} provides an overview of our Monte Carlo simulation, and how we use its outputs to compute physical fluxes. In Section~\ref{sec:binary_ejecta}, we present the numerically computed ejection spectra for various binary systems, and illustrate how black hole binaries are likely the most effective systems in the Galaxy at boosting DM particles. We then focus on this class of binary, and estimate the corresponding flux from a few representative populations within the Milky Way in Section~\ref{sec:bin_pop}. We convert the fluxes to velocity distributions in Section~\ref{sec:vel_dist}. From these, we estimate the extended mass sensitivity of LZ and PandaX-4T, and discuss the prospects for other detectors, in Section~\ref{sec:DD_sensitivity}. We conclude and provide an outlook for future analyses in Section~\ref{sec:outlook}. Several Appendices provide further details of the simulation and the resulting distributions of DM ejecta, including sample DM trajectories, the impact of binary eccentricity, mass ratio and barycenter motion. 

Unless otherwise specified, we work in natural units where $\hbar = c = 1$ and $G = M^{-2}_{\rm planck}$. The gravitational ejection energies quoted throughout this work are per unit mass. However, nuclear recoil energies in Sec.~\ref{sec:DD_sensitivity} are dimensionful.

\section{Gravitational Ejection}
\label{sec:grav_ejection_analytic}
Before analyzing the simulated ejection spectra, it is convenient to first gain an analytic understanding of the ejection mechanism and the maximum energy that can be obtained from it. We will, however, limit ourselves to the restricted 3-body problem, $i.e.$ $m_\chi \ll M_2 \ll M_1$ where $M_1$ is the mass of the heavy primary object in the binary, $M_2$ is the mass of the lighter secondary companion orbiting it, and $m_\chi$ is the mass of the incoming DM particle. In this regime, the resulting boost can be directly related to the gravitational deflection of the DM by the secondary \cite{2023EJPh...44b5002B}. 

\begin{figure*}[t!]
    \centering
    \hspace*{0.3cm}
    \includegraphics[width=0.8\textwidth]{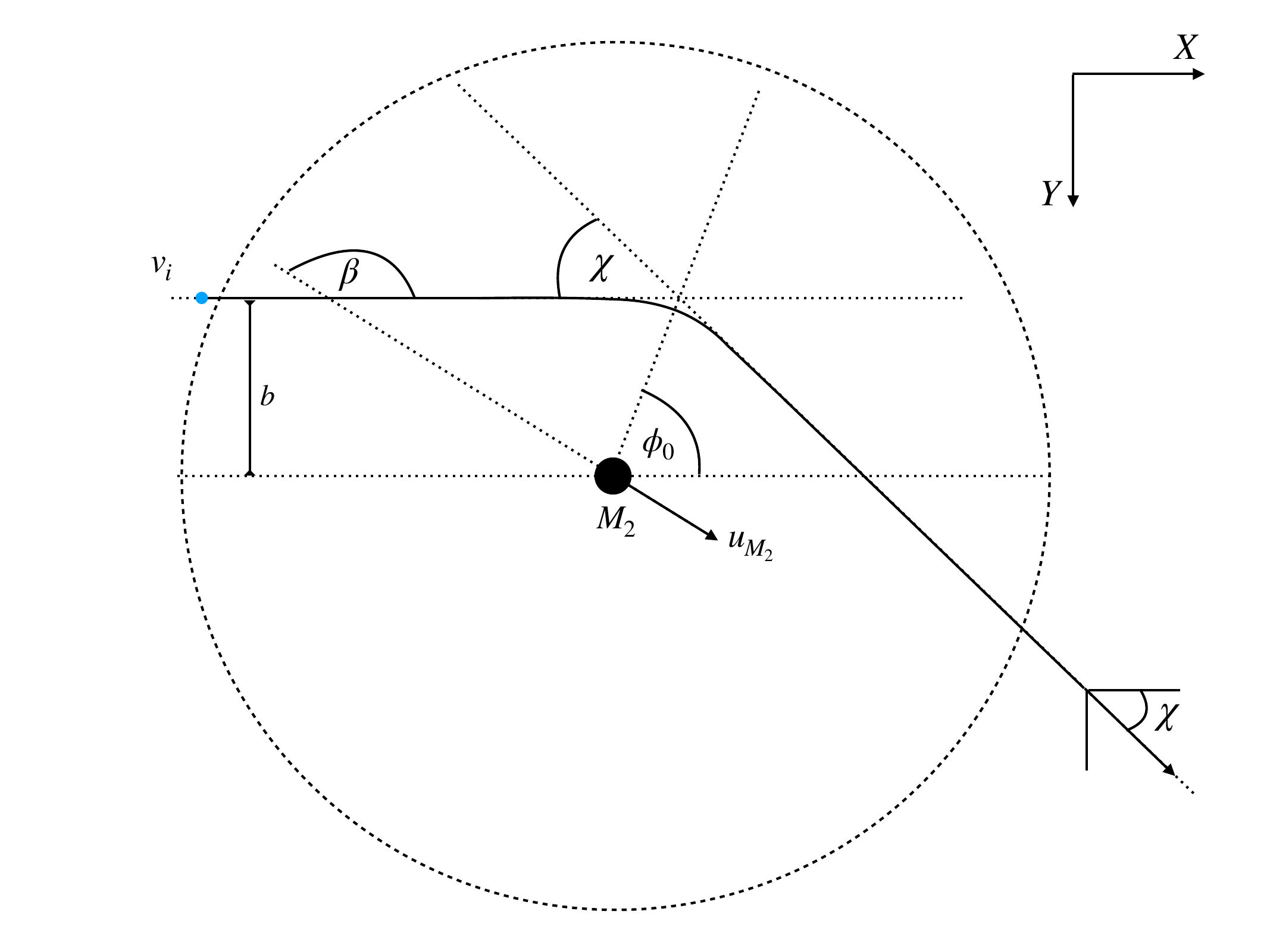}
    \caption{Schematic diagram of the deflection process for a DM particle in the rest frame of the binary's secondary companion (see text for details).}
    \label{fig:slingshot_schem}
\end{figure*}

Figure~\ref{fig:slingshot_schem} illustrates the process for an arbitrary scattering angle $\chi$ in the rest frame of the secondary, starting from an initial speed $v_i$ and impact parameter $b$ in this frame. Note that, in the primary's frame, the speed of approach is approximately $\sqrt{v^2_{1, \rm esc}(R_{\rm orb}) + v_\infty^2}$ within the restricted 3-body problem, where $v_{1, \rm esc}(R_{\rm orb})$ is the primary's escape velocity at the secondary's position, and $v_\infty$ is the initial halo velocity at infinity. The deflection is assumed to occur close enough to the secondary so that its gravitational potential is fully dominant relative to the primary; this sphere of influence is denoted by the dashed circle. The secondary's velocity, as measured in the rest frame of the primary, is also indicated, and it is assumed to form an angle $\beta$ with the negative x-direction in the secondary's frame. At the point of closest approach $r = r_{\rm min}$, the polar angle $\phi_0$ is \cite{landau1982mechanics}
\begin{equation}
    \phi_0 = \int^{\infty}_{r_{\rm min}} \frac{(b/r^2) \, dr}{\sqrt{1 - (b/r)^2 - 2 GM_2/(v^2_i \, r)}} = \arccos\left(\frac{x}{\sqrt{1+x^2}}\right)~,
\end{equation}
where $x = GM_2 / (b\, v^2_i)$. 

The scattering angle $\chi$ is related to $\phi_0$ via 
\begin{equation}
    \chi = \pi - 2\phi_0~,
\end{equation}
since the trajectory is symmetric about the point of closest approach.
Combining the identities $\arccos\left(1/\sqrt{1+z^2}\right) = \arctan(z)$ and $\pi/2 - \arctan(z) = \rm arccot(z)$, its resulting value can be written in terms of the speed of approach, impact parameter, and mass of the secondary, 
\begin{equation}
    \tan\left(\frac{\chi}{2}\right) = \frac{G M_2}{b \, v_i^2}~.
    \label{eq:scattering_angle}
\end{equation}
Let us first consider the particle motion in the frame indicated in the upper right of Fig.~\ref{fig:slingshot_schem}.
Neglecting any energy exchange for now between the incoming particle and the binary companion, the initial velocity vector of the DM is 
\begin{equation}
    \mathbf{v}_i = v_{i}  \, (1,0)~,
\end{equation} 
whereas the final velocity vector after scattering is simply
\begin{equation}
    \mathbf{v}_f = v_{i} \, (\cos\chi, \sin\chi)~.
\end{equation}
Before boosting to the primary's rest frame, it is convenient to rotate the axes so that the X-direction above is aligned with the companion's motion. This is parameterized by the angle $\beta$, and is a measure of whether the particle approaches from the leading ($\beta < 90^\circ$) or trailing ($\beta > 90^\circ$) side of the companion's motion. Upon performing this rotation, both initial and final vectors are then
\begin{equation}
    \mathbf{v}_i \rightarrow v_{i}\,(- \cos\beta, -\sin\beta)~,
\end{equation}
\begin{equation}
    \mathbf{v}_f \rightarrow v_{i}\,(- \cos(\beta+\chi), \, -\sin(\beta+\chi))~.
\end{equation}
An additional minus sign is introduced since we have defined the angle $\beta$ so that $\beta = 0^\circ$ when the velocity of the particle is antiparallel to the companion's motion. Boosting now to the primary's frame, the initial and final velocity vectors become
\begin{equation}
    \mathbf{u}_i = (u_{M_2} - v_{i}\cos\beta, -v_{i} \sin\beta)~,
\end{equation}
\begin{equation}
    \mathbf{u}_f = (u_{M_2} - v_{i}\cos(\beta+\chi) , -v_{i} \sin(\beta+\chi))~.
\end{equation}

Comparing the initial and final state velocities in the primary's frame, it can immediately be seen that the DM either gained or lost energy depending on the combination of angles $\beta$ and $\chi$. Since the scattering is elastic in the companion’s frame, the energy change arises solely from the Galilean boost back to the primary frame. This energy change, normalized to DM mass, is given by
\begin{equation}
    \Delta \varepsilon = u_{M_2} v_{i} \, \left(\cos\beta - \cos(\beta+\chi)\right)~.
    \label{eq:delta_epsilon}
\end{equation}
Eq.~\eqref{eq:delta_epsilon} indicates that the DM gains energy whenever the scattering increases its projected velocity along the companion’s direction of motion. For example, for angles $0^\circ <\beta < 90^\circ$, any scattering angle $0^\circ < \chi < 180^\circ$ results in a net energy gain, since the scattering process will increase the forward velocity component. If $\beta > 90^\circ$, there is an upper bound on the scattering angle that achieves energy gain, as the velocity vector can now be deflected in the opposite direction to the companion's motion if $\chi$ is too large. Eq.~\eqref{eq:delta_epsilon} also indicates that the maximal energy gain is $2 \, u_{M_2} v_i$, which occurs when $\beta = 0^\circ$ and $\chi = 180^\circ$. In this ``slingshot'' case, an incoming particle oppositely aligned to the companion's motion is now maximally deflected into a direction completely aligned with the companion's motion. Consequently, systems in which the companion is orbiting within a deep gravitational potential will maximize $\Delta \varepsilon$, since both $v_{M_2}$ and $v_i$ can potentially be very large. Although this may appear to violate energy conservation, in the scattering process between the DM and the companion, we have neglected the energy lost or absorbed by $M_2$, which scales as $m_\chi / M_2$ and is therefore negligible. In Appendix~\ref{app:DM_orbits}, we show a sample of trajectories with net energy-gain directly obtained from our simulation below for a fiducial, short-period double black hole binary. 

It is important to emphasize that Eq.~\eqref{eq:delta_epsilon} is technically only valid for the restricted 3-body problem, in which the heavy component is effectively at rest and a sufficiently close encounter for the DM can be described as a two-body hyperbolic scattering event in the rest frame of the secondary. While this provides useful intuition for how the energy gain scales with orbital speed and deflection angle, we caution against its limitations when applied to more general systems. In particular, for the most compact equal-mass black hole binaries we focus on, we have numerically observed ejections that exceed this estimate by an order unity factor. In this case, the gravitational spheres of influence of the two companions overlap significantly, leading in some cases to more complex gravitational interactions that cannot be described in terms of a single hyperbolic encounter. Furthermore, this calculation does not provide information about the probability distribution of the ejecta when given some initial velocity distribution of the incoming particles. For these reasons, a fully numerical treatment of the problem is required, which we develop below.

\section{Simulation}
\label{sec:sim_description}
Our simulation evolves in parallel an ensemble of 3-body systems consisting of the primary and secondary stellar objects, plus a third particle representing the DM. For each individual realization, it tracks the orbital evolution of all particles until eventual ejection or capture of the DM is reached, where these outcomes are defined more precisely below. Based on the ensemble, statistics are then drawn on the probability of ejection with a certain energy and direction. The only underlying assumptions of this method are the following:
\begin{enumerate}
    \item We restrict ourselves to the regime where the DM is much lighter than either binary component ($m_\chi \ll M_2$), and therefore there is no back-reaction on the binary dynamics. 
    \item We assume that the DM is heavy enough to be treated as particle-like ($1/m_\chi \ll$ all distance scales in the binary). Although this assumption has no practical implications for the simulation scheme, it limits the mass range for which the boosted particle flux computed with this framework is valid. Since we are mainly interested in the MeV to GeV scale, this condition is always satisfied here. 
    \item We further assume that DM has no sizable self-interactions, which would technically require an N-body simulation to accurately track its orbital evolution.
\end{enumerate}
Under such assumptions, the trajectory of each DM particle is therefore solely determined by the time-dependent gravitational potential of the binary components, justifying the ensemble approach. 

In each realization, the stellar objects are initialized in the binary's barycentric ($i.e.$ center of mass) rest frame, with initial conditions drawn to exactly reproduce the Keplerian solution of the isolated two-body problem. In other words, their initial positions and velocities are fixed by the preset primary and secondary masses $M_{1,2}$, period $P$ and eccentricity $e$ such that the stars follow the corresponding Keplerian binary orbit. The binary's true anomaly $\psi$ is randomly drawn from a uniform distribution between $(0 , 2\pi)$. With both binary components initialized, the DM particle is then injected on a bounding sphere much larger than the binary's spatial extension, of order $(10 - 200) \times$ its semimajor axis $a$. Its initial velocity is drawn from a truncated Maxwell–Boltzmann distribution characterized by a velocity dispersion $\sigma_\chi$, a maximum velocity given by the Galactic escape value $v_{\rm Gal}$, and a bulk wind velocity $\textbf{V}_{\rm w}$. We impose cuts requiring the DM to have initial positive energy and its initial velocity to at least partially point toward the interior of the sphere. The orbital evolution of all particles is then obtained through a Velocity Verlet algorithm \cite{hairer2006geometric}. Because the DM particle is effectively massless, the binary components remain on their Keplerian orbits to excellent approximation, while the DM particle follows a chaotic trajectory that may result in capture or ejection. We classify the particle as \textit{captured} if it crosses the physical radius of either stellar object, or else has a negative energy upon reaching the maximum allotted time\footnote{A detailed study of capture dynamics in binaries is left for future work. In the present analysis, “capture” should be understood as removal from the simulation.}. Conversely, it is classified as \textit{ejected} if it exits the simulation boundary at any time with positive energy. In the latter case, the final energy and ejection direction of the DM particle are subsequently recorded and binned.

We are interested in the differential ejection rate with respect to energy and direction, as this output ultimately determines the terrestrial flux. We estimate it from combining the probability of ejection at a given energy and direction, obtained from the simulation, with the physical rate of particles that would be interacting with the binary,
\begin{equation}
   \frac{d^3N_\chi}{d\Omega \, d\varepsilon \, dt} \simeq \mathcal{F}_{\rm sim} \times \frac{d^2p}{d\Omega \, d\varepsilon}~.
    \label{eq:diff_spec_gen}
\end{equation}
Above, $\Omega$ is the solid angle and $\varepsilon$ is the ejection energy. On the right hand side, the flux factor $\mathcal{F}_{\rm sim}$ is given by
\begin{equation}
    \mathcal{F}_{\rm sim} \simeq \sqrt{\frac{2}{\pi}} \, \pi R_{\rm sim}^2 \, n_\chi \, \left[2\sigma_\chi \exp\left(-\frac{V^2_{\rm w}}{2 \sigma_\chi^2}\right) + \sqrt{2\pi} \left(\frac{V^2_{\rm w} + \sigma_\chi^2}{V_{\rm w}}\right) \, {\rm erf} \left(\frac{V_{\rm w}}{\sqrt{2} \, \sigma_\chi}\right)\right]~.
\end{equation}
where $n_\chi = \rho_\chi/m_\chi$ is the background number density of DM particles and $R_{\rm sim}$ is the spherical simulation boundary radius. The joint probability function is the primary simulation output, and is a function of all the internal parameters of the binary and the distribution of incoming particles at infinity, 
\begin{equation}
    \frac{d^2p}{d\Omega \, d\varepsilon} = \frac{d^2p}{d\Omega \, d\varepsilon}(P, e, M_1, M_2, \sigma_\chi, v_{\rm Gal} ,\mathbf{V}_{\rm w})~.
    \label{eq:bin_dependence}
\end{equation}
For all the systems we have simulated, we find the fraction of captured particles to be negligible compared to the ejection fraction by a few orders of magnitude. This means the total ensemble size is, to an excellent approximation, also the number of ejected particles. Therefore, we need not multiply Eq.~\eqref{eq:diff_spec_gen} by a global probability of ejection, as the capture outcome is extremely rare. While specific the ensemble size varies depending on the binary properties and statistical error tolerance, in all of our runs it ranged from $\sim 5 \times 10^7$ up to $\sim 10^{10}$.

It is important to note that the above procedure to compute the ejection spectrum, as given by Eq.~\eqref{eq:diff_spec_gen}, produces a two-component curve: a low-energy tail comprised of all the particles that never approach the binary too closely, and a high-energy tail comprised of all the particles that meaningfully interact with the binary's internal structure. The former is formally divergent in the limit that the simulation boundary becomes infinite, whereas the latter always converges to the same value regardless of the boundary size. This is because, at long distances, all particles undergo weak scatterings with the binary's nearly static monopole potential, forcing the simulation to always record an $\mathcal{O}(1)$ fraction of all simulated particles as ejected. As these soft gravitational scatterings occur for all impact parameter values up to the simulation boundary itself, the ejection rate associated with these events monotonically grows with $R_{\rm sim}$. By contrast, for hard ejections the relevant impact parameter range is bounded by the binary's intrinsic spatial extent, and the corresponding rate is therefore independent of $R_{\rm sim}$. We have tested the convergence of the high energy tail for several simulation volumes, confirming that the physically-relevant part of the spectrum is independent of the boundary size. Furthermore, we have also tested imposing a threshold energy gain for a particle to be recorded as ejected ($e.g.$ $\varepsilon_f \geq 1.1 \,\varepsilon_i$), and found that this procedure effectively removes all soft-scatterings from the counts. In this case, as expected, the spectrum converges for all energies. From here on, for simplicity, we will refer to the well-defined high-energy tail of interest as the ejection spectrum.

In principle, both $\Omega$ and $\varepsilon$ are not statistically independent variables. However, as we justify in Appendix~\ref{app:ejecta_dist_approx}, as a reasonable first approximation we factorize the joint probability, 
\begin{equation}
    \frac{d^2 p}{d\Omega \, d\varepsilon} \simeq \frac{dp}{d\Omega} \times \frac{dp}{d\varepsilon},
    \label{eq:factorize_energy_angle}
\end{equation}
where each probability distribution is numerically computed as 
\begin{equation}
    \frac{dp}{d\Omega} \rightarrow \frac{1}{\Delta \Omega_{i}} \left(\frac{N(\theta_i,\phi_j)}{N_{\rm ej}}\right)~,
\end{equation}
\begin{equation}
    \frac{dp}{d\varepsilon} \rightarrow \frac{1}{\Delta \varepsilon} \left(\frac{N(\varepsilon_i)}{N_{\rm ej}}\right)~.
\end{equation}
Above, $N(\theta_i, \phi_j)$ and $N(\varepsilon_i)$ respectively are the number of ejected particles with direction and energy between $(\theta_i, \ \theta_i + \Delta\theta)$, $(\phi_j,\ \phi_j + \Delta\phi)$, and $(\varepsilon_i, \varepsilon_i + \Delta \varepsilon)$. We use standard spherical coordinates, with the binary orbital plane lying at $\theta = \pi/2$, and barycenter placed at the coordinate origin. The total number of particles that were ejected in the simulated ensemble is denoted by $N_{\rm ej}$. Obtaining the differential spectrum without marginalizing over directions would require significantly more computational resources than currently used here. As a first exploratory study, we have made the simplifying assumption that the energy and direction probabilities can be factorized.

To further simplify calculations, we will also approximate the angular distribution of the ejecta as isotropic by fixing 
\begin{equation}
    \frac{dp}{d\Omega} \simeq \frac{1}{4 \pi}~.
    \label{eq:isotropic_prob}
\end{equation}
This approximation is also justified in Appendix~\ref{app:ejecta_dist_approx}. While ejections preferentially occur close to the binary’s orbital plane, we find this anisotropy effect is generally mild. Moreover, binaries tend to have their orientations randomized both at formation \cite{2016ApJ...827L..11O} as well as over time from the constant interactions with field stars \cite{Merritt:2001dc}. Since we are interested in the collective contribution from binaries of the Milky Way, we expect the isotropic approximation to average out and therefore be appropriate for our purposes. 

Finally, we remark that our simulation only assumes Newtonian gravity. When either binary component is a black hole, we impose a cutoff distance of $10\times$ its gravitational radius. Particles that, at any point, approach the black hole closer than this cutoff are removed from the counts for added safety. However, as previously stated, this capture outcome is extremely infrequent.

\section{Simulated Binaries in the Milky Way}
\label{sec:binary_ejecta}
We now proceed to compare the numerically obtained ejection spectra for various existing binary systems, and validate the simple analytic picture of Sec.~\ref{sec:grav_ejection_analytic}. In particular, we show how double black hole binaries are likely the most effective systems for this ejection process. 

\begin{figure*}[t!]
    \centering
    \hspace*{0.7cm}
    \includegraphics[width=\textwidth]{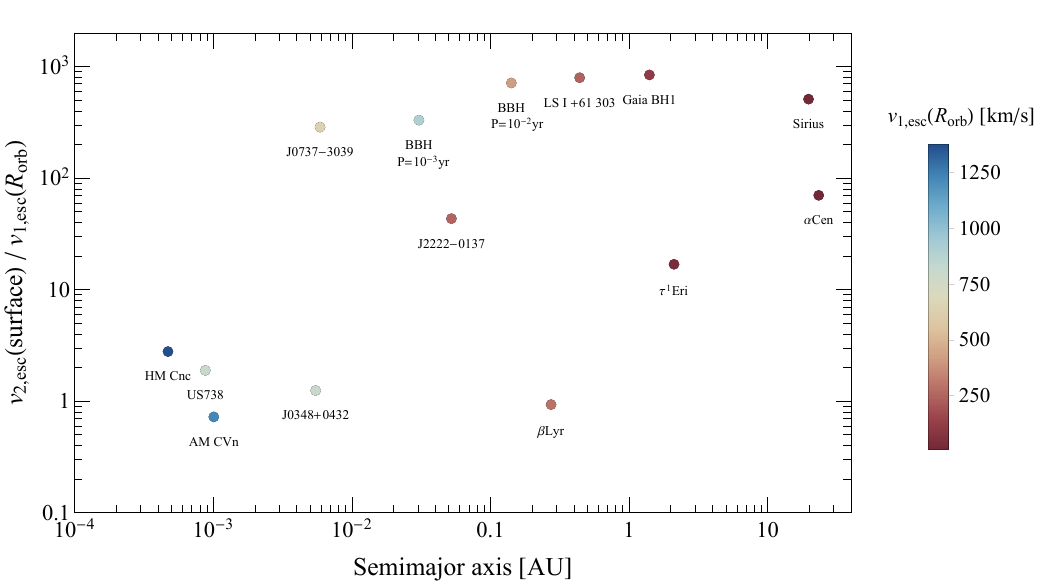}
    \caption{A sample of observed (or representative hypothetical) binary systems in the Milky Way, classified by semimajor axis. Masses are labeled such that $M_2 \leq M_1$. The ratio of the secondary's escape velocity to the primary's at the orbital distance is related to the deflection probability, while the primary’s escape velocity, shown by the color scale, determines the maximum ejection energy. The circular, equal-mass black hole binaries we focus on in this work are indicated by the label `BBH',  in both cases with total mass $M_1 + M_2 = 28 \, M_\odot$.}
    \label{fig:bin_sys_comp_1}
\end{figure*}

Let us first consider the dependence of the spectrum on the binary's separation and escape velocity for an idealized system with two point masses. For larger orbital separations (and correspondingly longer periods for fixed total mass), the effective impact parameter for transiting DM particles to interact with the binary increases, resulting in a larger total rate of ejected particles. However, the binary's gravitational well becomes shallower, which reduces the orbital velocity of each companion, as well as the initial speed of the DM during the encounter, therefore suppressing the maximum energy that can be gained, $cf.$ Eq.~\eqref{eq:delta_epsilon}. In other words, increasing the binary size increases the total rate but softens the ejection spectrum, and vice versa.  

Turning now to real binaries, the optimal systems must therefore achieve a balance between sourcing a deep gravitational well (from either one or both binary components) and avoiding too small orbital separations. The probability for each component to produce large-angle, high-energy deflections of transiting particles must also be high, which in turn requires encounters at small distances of closest approach where the gravitational field is strongest, $cf.$ Eq.~\eqref{eq:scattering_angle}. These requirements are sufficiently restrictive for us to focus on a single type of object. Binaries containing at least one stellar or planetary component ($e.g.$ WZ Sge \cite{1998PASP..110.1132P, Steeghs:2007ta}, $\beta$ Lyrae \cite{Zhao:2008sw}, LS I $+61 \ 303$ \cite{MAGIC:2006pue}, Gaia BH1 \cite{2023MNRAS.518.1057E}) cannot generally produce a hard ejection spectrum. When one component is extended, there is an inherent lower bound to the orbital size based on its tidal disruption radius. As a result, such systems in nature typically have long periods for which deflections occur in a shallow gravitational field and energy gains are minimal for the purposes of this analysis. Moreover, these objects source relatively weak gravitational fields outside their volume, and their hard surface imposes a lower cutoff on the distance of closest approach, strongly suppressing the probability for high-energy deflections. It then follows that binaries composed of either white dwarfs, neutron stars, or black holes will be more promising candidates. Owing to their compactness, binaries comprised of these objects can have much smaller orbital distances while also allowing for a higher deflection probability. Among these, there exist for instance ultracompact double white dwarf binaries ($e.g.$ AM CVn \cite{2023AcA....73..227S}, US738 \cite{2014MNRAS.444L...1K}, HM Cancri \cite{Green:2025qbw}) with periods of order minutes. In these systems, the white dwarfs orbit so closely that deflections can achieve significant energy gains, as these occur in a region of much higher escape velocity. However, such short period values combined with the relatively small white dwarf masses results in a small ejection rate. While the deflection probability is somewhat increased for systems containing a neutron star ($e.g.$ PSR J1946+2052 \cite{2018ApJ...854L..22S}, J0348+0432 \cite{Antoniadis:2013pzd}, J0737$-$3039 \cite{2003Natur.426..531B}, J2222$-$0137 \cite{Guo:2021bqa}), the same qualitative conclusion about the ejection rate is reached given the rough similarity in mass between neutron stars and white dwarfs. The remaining possibility is black holes produced by stellar progenitors. In contrast to the previous cases, these have typical masses $(10 - 20)\times$ larger than every other object, allowing binaries to have a much larger physical size without paying a large penalty in terms of the depth of the binary's gravitational potential. Furthermore, their extreme compactness permits very close encounters, maximizing the probability for high-energy deflections.

\begin{figure*}[t!]
    \centering
    \includegraphics[width=0.9\textwidth]{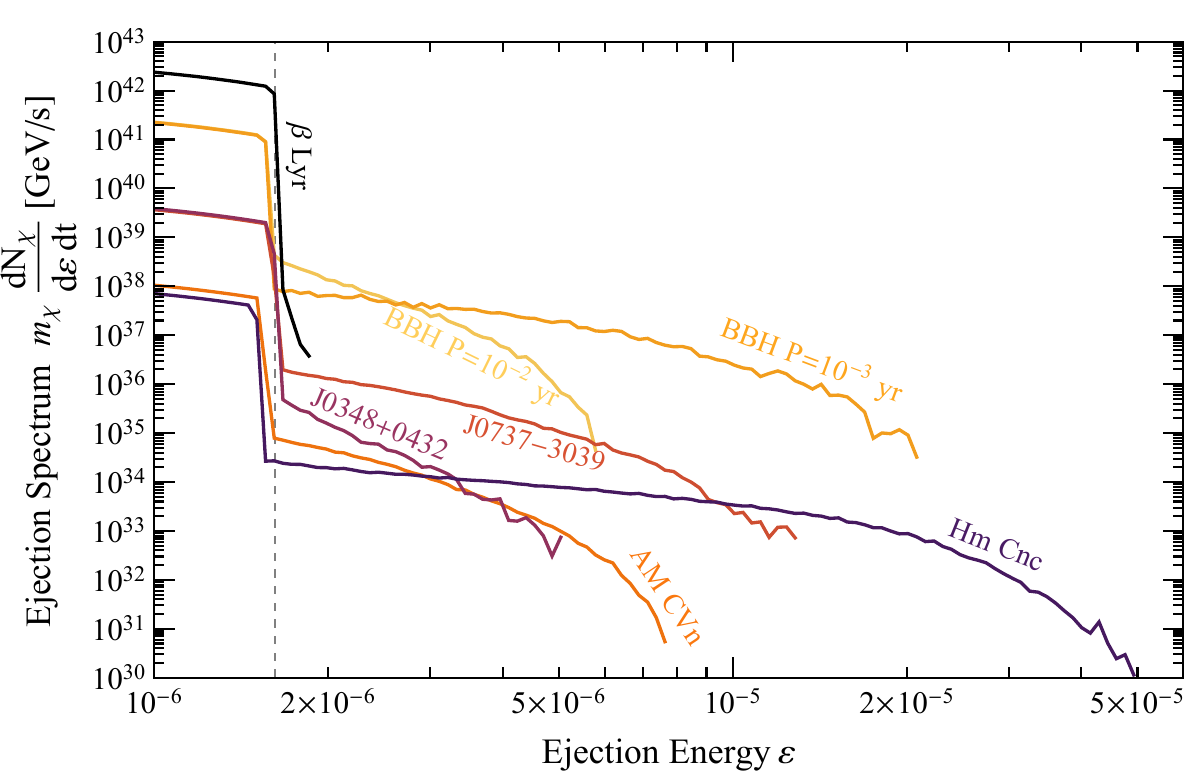}
    \caption{Simulated ejection spectra for a sample of the systems shown in Fig.~\ref{fig:bin_sys_comp_1}. For ease of comparison, in all cases we assume a background DM density of $\rho_\chi = 0.42 \ \rm GeV \, cm^{-3}$, velocity dispersion $\sigma_\chi = 240 \ \rm km/s$ and Galactic escape velocity cutoff $v_{\rm Gal} = 546 \ \rm km/s$. The latter is indicated in terms of energy by the vertical dashed line. Note that the ejection energy in the horizontal axis is normalized to DM mass.}
    \label{fig:bin_sys_comp_2}
\end{figure*}

Figure~\ref{fig:bin_sys_comp_1} shows a sample of observed binary systems, together with the double black hole binaries we will focus on later. In all cases, the indices for each object are defined such that $M_2 \leq M_1$. Each system is classified by semimajor axis on the horizontal axis, and by the ratio of the secondary’s surface escape velocity to the primary’s escape velocity at the orbital distance on the vertical axis. The color scale indicates the escape velocity of the primary, also at the secondary's position. Based on the above arguments, these quantities ultimately determine the ejection spectrum: the orbital separation determines the characteristic impact parameter for particles to interact with the binary, and therefore sets the overall flux of ejected particles. The ratio of escape velocities determines the effectiveness for the secondary at deflecting incoming particles, setting the slope and normalization of the ejection spectrum. The escape velocity of the primary determines the scale for energy gain, as this sets the orbital velocity of the secondary and, if large enough, the DM's velocity at the encounter point ($cf.$ Eq.~\eqref{eq:delta_epsilon}). It can be seen that binaries containing extended objects predominantly lie at large orbital separation values, where the ratio of escape velocities is large because the primary's gravitational field is weak at such distances. For these systems, energy gain through gravitational interactions is severely limited given the small orbital velocities involved. 
Binaries comprised of dual compact objects, by contrast, can present much shorter orbital separations without disruption, implying larger orbital speeds and therefore stronger energy gains from deflections. Double black hole binaries, in particular, can present both large orbital distances while retaining a relatively steep gravitational well, owing to their much larger masses relative to any other object. Given their maximal escape velocity, they are also favored in terms of deflection efficiency by their large ratio of escape velocities. Systems comprised of neutron stars and white dwarfs have, by contrast, smaller orbital separations, with ultracompact white dwarf binaries mostly populating the smallest orbital sizes. While these systems have deep potential wells which kinematically permit large energy gains, their deflection probability is lower, as the typical escape velocity of these objects can be weaker compared to their orbital velocity. Note that we have not included in this figure binaries including $e.g.$ supermassive black holes, which can present large orbital separations without penalty on the secondary's orbital velocity or deflection efficiency, if the latter is sufficiently compact. An example of this case is discussed in Sec.~\ref{sec:bin_pop}.

Figure~\ref{fig:bin_sys_comp_2} shows the numerically computed ejection spectra for a subsample of systems shown in Fig.~\ref{fig:bin_sys_comp_1}, through our simulation procedure. In all cases, we have assumed the initial distribution parameters as those given by the Standard Halo Model (SHM), including the truncation at the Galactic escape velocity denoted by the vertical dashed line. As described in Sec.~\ref{sec:sim_description}, the numerical ejection spectra have two components: a low-energy tail arising from weak, distant gravitational encounters that produce only small velocity perturbations, and a high-energy tail produced from binary deflections at energies roughly above the Galactic escape velocity. We have included the former for completeness; however, its overall normalization depends on the simulation boundary size and therefore does not constitute a robust physical prediction to the left of the vertical line. 
The spectra broadly match the above expectations: binaries containing main-sequence stars are grossly inefficient at ejecting particles at high energy. In fact, of the main-sequence systems listed in Fig.~\ref{fig:bin_sys_comp_1}, only $\beta$ Lyr is able to produce non-negligible ejecta above the Galactic cutoff. Ultracompact white dwarf binaries, such as AM CVn or HM Cnc, and neutron star binaries, like PSR J0348+0432 or J0737$-$3039, can achieve sizable boosts at the expense of a tiny flux. Compact black hole binaries reach about the same ejection energies without compromising the rate. This comparison confirms that the compactness of both the orbital configuration and individual companions are the key factors determining both the hardness and magnitude of the ejection spectrum. For these reasons, of all the possible systems in the Milky Way, double black hole binaries with short periods around $\lesssim 10^{-2} \ \rm yr$ are likely among the most effective sources of gravitationally-boosted DM. 

\section{Target Binary Populations}
\label{sec:bin_pop}
We now consider a few representative populations of double black hole binaries in the galaxy, and determine their ejection spectra as a function of their internal parameters. Specifically, we will consider three distinct populations: 1) binaries near the Sun's position, 2) binaries residing in the Milky Way's bulge, and 3) the black hole population in the nuclear star cluster at the Galactic Center. 

\subsection{Local Binary Black Holes}
\label{sec:local_2BH}
We begin by considering the population of local double black hole binaries. Their associated flux of boosted DM requires as inputs the relevant distributions for the internal binary parameters, as well as the spatial distribution of these systems. To obtain well-motivated estimates, we will utilize the synthetic catalogue presented in Ref.~\cite{2018MNRAS.480.2704L}, which combines a high-resolution cosmological simulation of a Milky Way-like galaxy within the FIRE suite with a binary population synthesis code. It contains a detailed dataset of binary position, period, and masses, among other properties. 

Simulating each predicted system in full is computationally unfeasible, given the already large amount ($\sim 600$) within $\lesssim 1 \ \rm kpc$ distance. We therefore adopt a set of simplifying assumptions that allow us to estimate the collective contribution from this population:
\begin{enumerate}
    \item We consider only circular binaries, $i.e.$ we neglect eccentricity. Eccentricity is in fact not listed in the catalogue entries, as it is argued in Ref.~\cite{2018MNRAS.480.2704L} that most binaries will be circularized to eccentricities $e \lesssim 0.15$ at present time. In Appendix~\ref{app:eccec}, we show the effects of finite eccentricity on the ejection spectra, but we note here that it has a small impact on the predictions unless the binary is extremely eccentric. 
    \item We consider equal-mass binaries, and use the average binary total mass predicted by the catalogue, $M_1 + M_2 \simeq 28 \, M_\odot$. In Appendix~\ref{app:mass_ratio}, we show the effects of a variable mass ratio on the ejection spectra. As with eccentricity, this parameter does not have a major effect on the predicted flux for this class of binaries derived from stellar progenitors (although see Appendix~\ref{app:M_squared_scaling} for extreme mass ratios). 
    \item The orbital period is the remaining internal parameter that significantly affects the ejection spectrum. Rather than simulating each system individually, we bin binaries by period and compute representative ejection spectra for each bin.
    \item We treat the contribution from local binaries as a diffuse flux, integrating along the line of sight over the binary spatial density. We restrict this integration to distances $D \lesssim D_{\rm max} = 1 \ \rm kpc$, beyond which the catalogue predicts a highly anisotropic distribution of black hole binaries. 
\end{enumerate}
Under these approximations, we express the resulting DM flux from local binary black holes as 
\begin{equation}
   \Phi_{\rm local} \simeq \int_{D< D_{\rm max}} \, n_{\rm BBH} \, \left(\frac{\mathcal{F}_{\rm sim}}{4\pi D^2}\right) \left(\sum_{P} \omega_{P} \, \frac{dp}{d\varepsilon}\right) \, D^2 \, dD \, d\Omega~,
   \label{eq:local_bbh_flux1}
\end{equation}
where $D$ is the line-of-sight distance, and the sum runs over all binned periods. We denote the fractional abundance in period by $\omega_P$, which we compute directly from the catalogue. Only a narrow range of binary periods actually contributes with particle energies much beyond the galactic escape cutoff. Periods beyond $\gtrsim 5 \times 10^{-2} \ \rm yr$ result in binaries producing ejecta too soft to meaningfully extend the sensitivity of direct detection experiments, whereas periods below $\lesssim 10^{-3} \ \rm yr$ produced hardened spectra but are extremely rare. For simplicity, we have performed a coarse-grained scan of this parameter where we consider period bins 
\begin{equation}
    P = \left\{0.075 , \ 0.10 , \ 0.25, \ 0.50, \ 0.75,  1.00, \ 2.50, \ 5.00 \right\} \times10^{-2} \ \rm yr~,
    \label{eq:period_bins}
\end{equation}
with corresponding abundances
\begin{equation}
    \omega_P = \left\{0.16, \ 0. , \ 0.49, \ 2.27, \ 2.27, \ 3.08, \ 13.33, 14.47  \right\} \times 10^{-2}~,
\end{equation}
which approximately accounts for the whole range of relevant values. In total, about $36\%$ of the local black hole binaries are contained in this period range.

As we integrate over a small local volume compared to the scale of the Milky Way, we further assume a fixed binary black hole number density $n_{\rm BBH}$. The resulting boosted DM flux from this population is then 
\begin{equation}
   \Phi_{\rm local} \simeq n_{\rm BBH} \, \sum_{P} \omega_{P} \int_0^{D_{\rm max}} \! \! \mathcal{F}_{\rm sim} \left(\sum_{P} \omega_{P} \, \frac{dp}{d\varepsilon}\right) \, dD~ \simeq n_{\rm BBH} \, D_{\rm max} \,\mathcal{F}_{\rm sim} \sum_{P} \omega_{P} \, \frac{dp}{d\varepsilon}
   \label{eq:local_bbh_flux2}
\end{equation} 
where we estimate $n_{\rm BBHs} \simeq \rm 1.47 \times 10^{-7} \, \rm pc^{-3}$ directly from the catalogue. 

\begin{figure*}[t!]
    \centering
    \includegraphics[width=0.9\textwidth]{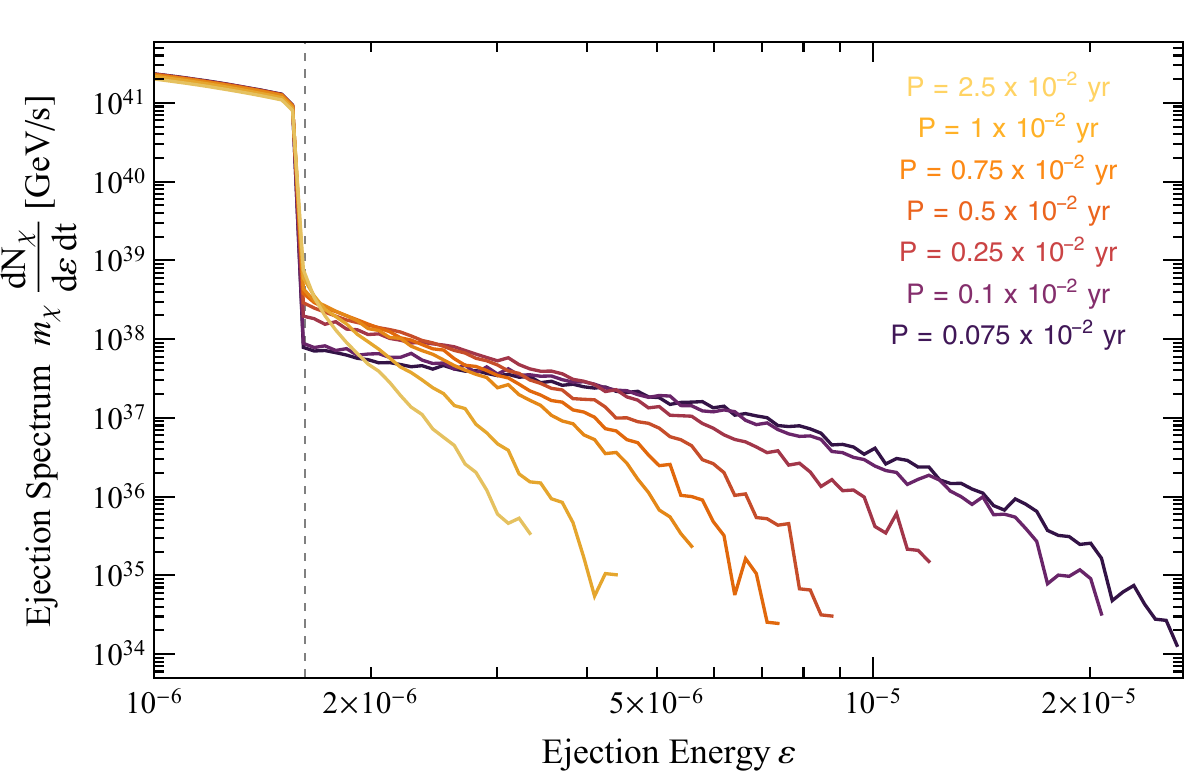}
    \caption{Simulated ejection spectra from equal-mass, circular black hole binaries with total mass $M_{1} + M_{2} = 28 \, M_\odot$. Each curve corresponds to a different binary period, as specified by the colors. These assume a background DM density of $\rho_\chi = 0.42 \ \rm GeV \, cm^{-3}$, velocity dispersion $\sigma_\chi = 240 \ \rm km/s$ and Galactic escape velocity cutoff $v_{\rm Gal} = 546 \ \rm km/s$. The vertical dashed line indicates the latter converted to energy. Note that the ejection energy in the horizontal axis is normalized to DM mass.}
    \label{fig:BBH_Spectra}
\end{figure*}

Figure~\ref{fig:BBH_Spectra} shows the mass ejection spectra that enter Eq.~\eqref{eq:local_bbh_flux2} for each period bin (excluding the fractional abundance factors $\omega_P$ above). As we are integrating over a small local volume, we have fixed the background DM density to that of the local position, $\rho_\chi = 0.42 \ \rm GeV \, cm^{-3}$ \cite{2024MNRAS.528..693O}, and adopted a velocity dispersion $\sigma_\chi = 240 \ \rm km/s$ and Galactic escape velocity cutoff $v_{\rm Gal} = 546 \ \rm km/s$ based on the best-fit values of Ref.~\cite{Folsom:2025lly}. Black hole binaries within this period range are capable of producing a flux of boosted particles with numerically observed ejection energies up to $\varepsilon \sim 2.75 \times 10^{-5}$, translating into maximum ejection speeds of approximately $\sim 2220 \ \rm km/s$ in their barycenter frame. For completeness, we also show the soft-scattering component of the spectra, which appears at energies approximately below the halo cutoff. As discussed in Sec.~\ref{sec:sim_description}, this is an artifact of the simulation that counts an unbounded number of weak, long-distance gravitational encounters for every particle simulated. The spectra converge to the same value in the low-energy regime because we have chosen the same boundary size in all the runs here. 

It is worth noting that alternative catalogues are available for this analysis. For instance, in Ref.~\cite{Olejak:2019pln} a synthetic black hole database was constructed using the population synthesis code StarTrack. This catalogue predicts a considerably larger number of black hole binaries in the galaxy, as well as a higher number density near the Solar System's position. This is driven by differences in the assumed metallicity of the progenitors, which can have a significant impact on the predicted number of compact objects formed. Consequently, this catalogue predicts a stronger binary-boosted flux relative to the results presented in this work. While we remain agnostic about the true abundance of black hole binaries in the galaxy, we employ the FIRE-based catalogue to remain conservative in our estimates.

\subsection{Bulge Binary Black Holes}
We next consider the binary black hole population residing in the bulge of the Milky Way. As in the local analysis, we rely on the FIRE-based synthetic catalogue \cite{2018MNRAS.480.2704L}, and adopt the same prior assumptions for the binary eccentricity, mass-ratio, and period distributions. 

For simplicity, we will compute the contribution from all binary black holes within a kpc-sized spherical volume around the Milky Way's center. Since the distance to the galactic center is $D_{\rm GC} \simeq 8.2 \ \rm kpc$, this allows us to approximate the population as a point-like source of boosted DM particles. We now fix the distance $D = D_{\rm GC}$ in the flux factor, and integrate over the source volume to determine the full flux of binary-boosted particles,
\begin{equation}
   \Phi_{\rm bulge} = \int_{R < \rm kpc} n_{\rm BBH}  \, \left(\frac{\mathcal{F}_{\rm sim}}{4 \pi D^2_{\rm GC}} \right) \left(\sum_{P} \omega_{P} \,\frac{dp}{d\varepsilon}\right) \, dV~.
   \label{eq:full_flux_bulge}
\end{equation} 
where $R$ is the galactocentric distance. We scan over the same period bins ($cf.$ Eq.~\eqref{eq:period_bins}), with fractional abundances now given by 
\begin{equation}
    \omega_P = \left\{ 0.025 , \ 0.028 , \ 0.42 , \ 1.32 , \ 1.67 , \ 1.62 , \ 6.99 , \ 9.08 \right\} \times 10^{-2} ~.
\end{equation}
Similarly to the local binary population, only a small fraction of about $21 \%$ of the total black hole binaries predicted in this region ($\sim 7.5 \times 10^5$) will meaningfully contribute to the boosted DM flux.

We have numerically verified from the catalogue entries that the spatial distribution of binaries within this $\rm kpc$-sized volume is highly isotropic, and can be well fitted by a piecewise power-law in galactocentric distance $R$. On the other hand, the individual contribution from each binary depends on the DM density around it, and is therefore also a function of $R$. We parameterize this dependence with a Navarro-Frenk-White (NFW) halo profile of the form \cite{Navarro:1995iw,Navarro:1996gj}
\begin{equation}
    \rho_{\chi}(R)  = \dfrac{\rho_{\chi}^0}{\left({R}/{R_s}\right)^{\gamma} \left(1+{R}/{R_s}\right)^{3-\gamma}}~,
    \label{eq:NFW_profile}
\end{equation}
where $R_s \simeq 20 \ \rm kpc$ is the scale radius, and $\gamma$ is the slope. The normalization constant $\rho_{\chi}^0$ is fixed so that $\rho_\chi \simeq 0.42 \ \rm GeV \ cm^{-3}$ at the Sun's position. For the slope, we consider two extremal values $\gamma = 1$ to $\gamma = 1.5$ based on halo contraction analyses of the inner Galaxy~\cite{2011arXiv1108.5736G,DiCintio:2014xia}. The remaining inputs that depend on galactocentric distance are the velocity dispersion $\sigma_\chi$ and escape cutoff $v_{\rm Gal}$. However, both quantities are difficult to measure towards the Galactic Center. In the case of velocity dispersion, observations hint at a narrow range within the central kiloparsec of order $200 - 300 \ \rm km/s$ \cite{2013PASJ...65..118S}. For simplicity, we use a constant $\sigma_\chi = 240 \ \rm km/s$ as in the local analysis, having verified that such a narrow variation does not translate into observable changes in the ejection spectra. In the case of the escape velocity cutoff, with the exception of the central $\sim \rm pc$ (see below), this parameter also varies by approximately $\sim 200 \ \rm km/s$ compared to the local position across various Milky Way potential models \cite{2013A&A...549A.137I}. The variation of this parameter can also be neglected and, for consistency, we also neglect the associated gravitational potential difference between source and detector. These approximations allow us to avoid re-simulating each system for each set of values of $\sigma_\chi$ and $v_{\rm Gal}$. The associated spectra for individual binaries of this population can then be read off Fig.~\ref{fig:BBH_Spectra} appropriately rescaled by the DM density factor $\rho_\chi(R)/\rho_\chi(R = 8.2 \ \rm kpc)$ for a given distance $R$. Furthermore, the integrand in Eq.~\eqref{eq:full_flux_bulge} is spherically-symmetric, and the integration can be performed solely in terms of Galactocentric distance, with $dV = 4 \pi R^2 dR$, to obtain the full distribution from this binary population.

Finally, we point out a potential correction to this calculation, which we find to be negligible. In our analysis, we equate the instantaneous ejection rate of DM particles from a binary with the corresponding detection rate at Earth. However, because DM particles require a finite time to propagate from the source to the detector, this is only valid if the binary properties do not evolve appreciably over the propagation time. The lowest ejection velocity of interest is roughly given by the escape velocity cutoff used in direct detection; as previously quoted this is approximately $\sim 546 \ \rm km/s$. For a typical propagation distance of $\sim 8.2 \ \rm kpc$, this yields a travel time of order $\sim 15 \ \rm Myr$. To justify the use of steady-state ejection rates, this timescale must be short compared to the binary's orbital decay time. The lowest binary periods we consider are of order $P \sim 10^{-3} \ \rm yr$ or, in terms of semimajor axis, $a \sim 0.03 \ \rm AU$ for the $M_{1,2} = 14 \, M_\odot$ black hole masses we consider here. In this short-period and low-eccentricity regime, the orbital decay is dominated by gravitational radiation, with a merger timescale of order $a/\dot{a} \simeq (64 G^3 M_1^3/5 a^4)^{-1} \sim 392 \ \rm Myr$, much longer than the ejecta propagation time even for those particles ejected at velocities near the cutoff. Therefore, the instantaneous ejection rate at source may also be reliably used at detection. 

\subsection{Black Holes in the Nuclear Star Cluster}
As a final galactic source of gravitationally-boosted DM, we consider the black holes in the Milky Way's nuclear star cluster, $i.e.$ the dense concentration of stars and compact objects surrounding the supermassive black hole Sagittarius A$^{\star}$ (Sgr.~A$^\star$) at the center of the galaxy. This region potentially constitutes the dominant contribution to the binary-boosted flux, given its extreme DM density as well as large black hole abundance. For instance, the most recent analysis of the star formation history predicts $\sim 2.2 \times 10^4$ binary black holes and $\sim 2.5 \times 10^5$ single black holes residing within $\sim 1.5 \ \rm pc$ around Sgr.~A$^\star$ \cite{2023ApJ...944...79C}. These estimates are in fact larger than some earlier predictions \cite{Hopman:2006xn,Alexander:2008tq}, as they incorporated recent high metallicity observations of this region \cite{2015A&A...573A..14R,2015ApJ...809..143D}.

As a simplified approach appropriate for the scope of this work, we consider two representative classes of systems: double black hole binaries, and binaries composed of Sgr.~A$^\star$ and an orbiting stellar-mass black hole. However, obtaining complete estimates for DM ejecta due to both of these populations is computationally demanding as well as subject to several assumptions concerning the binary parameter distribution. We will therefore restrict ourselves to simplified order-of-magnitude estimates, and leave a detailed estimation of the boosted DM flux from this region for future investigation.

Let us first consider the population of double black hole binaries whose barycenter orbits around Sgr.~A$^\star$. Such systems cannot generally transfer sufficient energy to DM particles bound to Sgr.~A$^\star$ to unbind them, unless their orbital periods are extremely short to the point where general relativistic effects likely become important. Moreover, based on the prior populations we have analyzed, binaries with such ultrashort periods are generally expected to be rare. One may instead consider halo DM particles that transit the Galactic Center region and are already unbound from Sgr.~A$^\star$. Neglecting the halo kinetic energy far from this region, their speed upon reaching the binary is effectively set by the escape velocity from Sgr.~A$^\star$ at the binary’s location,
\begin{equation}
v_{\rm SgrA^\star}(R) \simeq 1900~{\rm km/s} \, \left(\frac{R}{10^{-2} \, {\rm pc}}\right)^{-1/2}~.
\label{eq:Vesc_Sgr}
\end{equation}
While this avoids the additional energy threshold for ejecting particles out of Sgr.~A$^\star$'s potential well, the ejection rate is highly suppressed because of the large velocities involved, as now the DM must approach either binary component at tiny distances to gain energy. For these reasons, despite their expected large abundance, black hole binaries in this region make a negligible contribution to the binary-boosted flux relative to the bulge population discussed above. We have confirmed this expectation with simulations in which the initial velocity distribution of the particles is replaced by a delta-like function centered at Eq.~\eqref{eq:Vesc_Sgr}, for various galactocentric distances $R$.

The alternative system to consider is the combination of Sgr.~A$^\star$ as the primary with a black hole bound to it as the secondary (the latter could itself be a compact binary, provided its internal separation is much smaller than its orbital distance to Sgr.~A$^\star$). If the companion orbits sufficiently close, we expect such a system to achieve the highest ejection energies given the huge escape velocity of Sgr.~A$^\star$. However, with our current simulation setup, the computational time required to obtain ejection spectra with low statistical error is orders of magnitude greater than in previous cases. There are two reasons for this increase, both of computational and physical origin: first, black holes in this region have periods reaching up to $\rm 10^4 \, yr$, consequently imposing a huge simulation boundary size in our current setup. This makes it more likely that the injected DM particles, given their randomized directions, will not meaningfully interact with the binary. Second, as in the case of double black hole binaries, the DM particles now have a characteristic speed determined by Sgr.~A$^\star$'s gravitational potential, drastically reducing the probability for ejection unless a sufficiently close encounter occurs. In both cases, this imposes a huge increase in the required ensemble size to achieve a well-resolved spectrum.

For these reasons, we instead focus on the ejection spectrum for a single $50 \, M_\odot$ black hole on a circular $P = 1 \ \rm yr$ orbit or, equivalently, a galactocentric distance $R \simeq 162 \ \rm AU$. Even for this simplified system, the numerical spectrum is cumbersome to compute. We obtain it by computing the spectrum for a larger companion mass $M_2$, and rescaling the result by $(50 M_\odot/M_2)^2$. This scaling law arises from the gravitational scattering cross-section: the approximate duration of a collision is $\delta t \sim l/v_\chi$, with $l$ some characteristic interaction length, whereas the acceleration experienced by a particle is of the order $a \sim G M_2 / l^2$. The change in the perpendicular component of the velocity is then $\delta v_\chi \sim a \times \delta t \sim G M_2/ (l \, v_\chi)$. For an interaction to cause a significant deflection, we require $\delta v_\chi \sim v_\chi$, or in terms of a cross-section, $\sigma \sim l^2 \sim (GM_2)^2 / v_\chi^4$. Utilizing this $M_2^2$ scaling relation allows us to improve computational efficiency by simulating an artificially heavier secondary with a much higher deflection probability. In Appendix~\ref{app:M_squared_scaling}, we show the numerical verification of this scaling law for a wide range of companion masses. We focus as before on unbound DM, which allows for a simplified estimate of its velocity dispersion from Liouville's theorem,
\begin{equation}
    \sigma^2_\chi(R) \simeq \sigma^2_\chi(R_0) \, \left[1 + \left(\frac{v_{\rm SgrA^\star}(R)}{\sigma_\chi(R_0)}\right)^2\right]~,
\end{equation}
where $R_0 = 1 \ \rm pc$ is a reference distance that we fix to roughly the radius of influence of Sgr.~A$^\star$. The use of Liouville's theorem as a first approximation is justified as we are assuming collisionless DM throughout this work. Similarly, for the unbound particle density we use 
\begin{equation}
    \rho_\chi(R) \simeq \rho_\chi(R_0) \, \sqrt{1 + \left(\frac{v_{\rm SgrA^\star}(R)}{\sigma_\chi(R_0)}\right)^2}~.
\end{equation}
We assume $\sigma_\chi(R_0) = 240 \ \rm km/s$ as before, and compute $\rho_\chi(R_0)$ from Eq.~\eqref{eq:NFW_profile}. While this chosen system is an idealized setup, it will serve to illustrate the potential gain in sensitivity for direct detection experiments (and, potentially, neutrino detectors) in a more complete analysis of this central region, to be pursued in upcoming work. 

\section{Velocity Distribution of Ejecta}
\label{sec:vel_dist}
To determine the prospects for DM direct detection, we must first relate the binary-boosted flux from each population to the actual velocity distribution used to compute recoil rates. Under the isotropic ejection approximation, the 3D velocity distribution from a \textit{single} binary system, as seen by an observer at rest relative to the halo, is formally given by 
\begin{align}
    f_{\rm BBH}(\mathbf{v_\chi}) &= \mathcal{N}^{-1}_{\rm BBH} \, \frac{dp}{dv_\chi}\, \delta^{(2)}\!\!\left(\Omega - \Omega_0\right) \ , 
    \label{eq:dist_single}
\end{align}
\begin{equation}
    \mathcal{N}_{\rm BBH} = \int d^3\mathbf{v_\chi} \,\frac{dp}{dv_\chi}\, \delta^{(2)}\!\!\left(\Omega - \Omega_0\right)~,
\end{equation}
and we transform the numerical probability $dp/d\varepsilon$ from ejection energy $\varepsilon$ to ejection velocity via $\varepsilon = v_\chi^2/2$. We emphasize that the numerical probability $dp/dv_\chi$ is a simulation output and is by default normalized so that $\int (dp/d\varepsilon) \, d\varepsilon = \int (dp/dv_\chi) \, dv_\chi = 1$. The normalization factor $\mathcal{N}_{\rm BBH}$ accounts for the integration Jacobian involved in three-dimensional velocity space. The delta function $\delta^{(2)}\!(\Omega - \Omega_0) = \delta(\phi - \phi_0) \, \delta(\cos\theta - \cos\theta_0)$ forces the velocity vector to be aligned with the direction of the binary-boosting source. Note that the simulation is performed in the binary's barycenter rest frame. Therefore, utilizing the numerical probability directly amounts to assuming the binary system is also at rest relative to the halo, $i.e.$ that $v_\chi$ is defined in the halo rest frame. As we detail in Appendix~\ref{app:syst_motion}, the impact of the relative motion between the binary and the halo on the ejecta energy and angular distribution only has a moderate impact and, as with other approximations we have made, we expect its effect to average out when integrating over a large population. 

Turning now to \textit{populations} of binary black holes, for local systems we assume that the resulting distribution becomes isotropic upon stacking the individual contributions for each direction. In this case, we have
\begin{equation}
    f_{\rm BBH}(\mathbf{v_\chi}) = \mathcal{N}^{-1}_{\rm BBH} \, \sum_P \omega_P \, \frac{dp}{dv_\chi}~,
\end{equation}
\begin{equation}
    \mathcal{N}_{\rm BBH} = \int 4 \pi v^2_\chi \, \left(\sum_P \omega_P \, \frac{dp}{dv_\chi}\right) \, dv_\chi\, .
\end{equation}
However, for the bulge, as well as the nuclear star cluster, the source is highly directional. In this case, the distribution is given by Eq.~\eqref{eq:dist_single} with the replacement  
\begin{equation}
    \frac{dp}{dv_\chi} \rightarrow \sum_P \omega_P \, \frac{dp}{dv_\chi}~ ,
\end{equation}
and by fixing the direction enforced by the delta function to that of the Galactic Center.

We additionally require the number density of boosted particles. For a single system, this is 
 \begin{equation}
    n^{\rm (BBH)}_\chi = \int_0^{\infty} \frac{1}{v_\chi} \left(\frac{d^3N_\chi}{dA \, dv_\chi \, dt}\right) \, dv_\chi~,
\end{equation}
where $dA$ is the area differential. It can be directly related to the simulated ejection spectrum via
\begin{equation}
    \frac{1}{v_\chi} \left(\frac{d^3N_\chi}{dA \, dv_\chi \, dt}\right) = \frac{1}{D^2} \, \left(\frac{d^3N_\chi}{d\Omega \, d\varepsilon \, dt}\right) = \frac{\mathcal{F}_{\rm sim}}{4 \pi D^2} \, \frac{dp}{d\varepsilon}~,
\end{equation}
since $d\varepsilon = v_\chi \, dv_\chi$. Using this relation, we obtain
\begin{equation}
     n^{\rm (BBH)}_\chi = \int_0^{\infty} \frac{\mathcal{F}_{\rm sim}}{4 \pi D^2} \, \, \frac{dp}{d\varepsilon} \, \frac{d \varepsilon}{\sqrt{2 \varepsilon}}~.
     \label{eq:n_chi_single}
\end{equation}
\begin{figure*}[t!]
    \centering
    \includegraphics[width=0.9\textwidth]{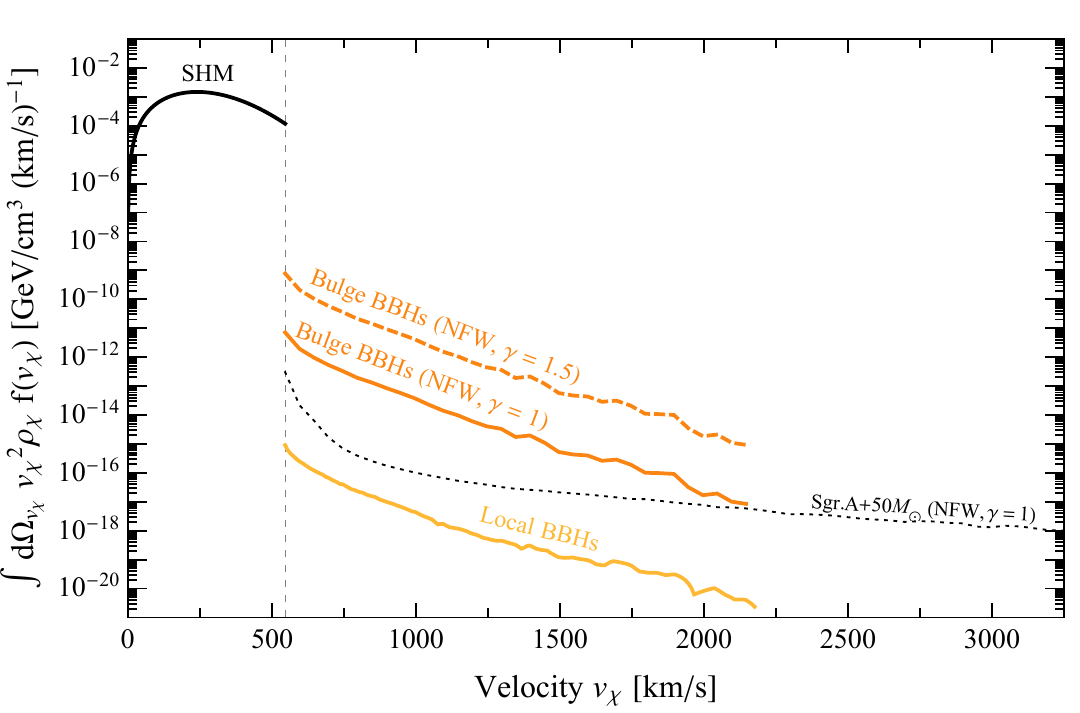}
    \caption{Binary-boosted DM velocity distributions from the local and bulge binary black hole populations, assuming all binaries are approximately at rest with respect to the Galactic halo. The Standard Halo Model distribution is shown for comparison, with the vertical dashed line indicating the local Galactic escape velocity. We also display the distribution from a single black hole orbiting Sgr.~A$^\star$ to illustrate the potentially dominant contribution from the Milky Way's nuclear star cluster.}
    \label{fig:Vel_Dist}
\end{figure*}
To determine the DM density from a \textit{population} of black hole binaries, it suffices to replace in Eq.~\eqref{eq:n_chi_single}
\begin{equation}
    \int_0^{\infty} \frac{\mathcal{F}_{\rm sim}}{4 \pi D^2} \, \frac{dp}{d\varepsilon} \, \frac{d \varepsilon}{\sqrt{2 \varepsilon}} \rightarrow \int_0^\infty \Phi_{\rm pop} \, \frac{d \varepsilon}{\sqrt{2 \varepsilon}}~,
\end{equation}
where $\Phi_{\rm pop} = \Phi_{\rm local}$ or $\Phi_{\rm bulge}$ are the fluxes computed previously. However, when combining the number density with the total velocity distribution, the sum over abundance factors $\omega_P$ should be understood to include the corresponding number density factor.

Figure~\ref{fig:Vel_Dist} shows the resulting velocity distributions from the black hole binaries outlined in Sec.~\ref{sec:bin_pop}, multiplied by mass density. For bulge binaries, we show the distribution for two choices of NFW slopes, in order to illustrate the potential reach spanned by the astrophysical uncertainty on this input. For comparison, we have included the SHM distribution, parameterized as a truncated Maxwellian distribution, 
\begin{equation}
    f_{\rm SHM}(\mathbf{v}_\chi) = \mathcal{N}^{-1}_{\rm SHM} \, \exp\left(-\frac{v_\chi^2}{\sigma_\chi^2}\right) \, \Theta(v_{\rm Gal} - v_\chi)~,
    \label{eq:vel_dist_SHM}
\end{equation}
where
\begin{equation}
    \mathcal{N}_{\rm SHM} = \, (\pi \sigma^2_\chi)^{3/2} \left[{\rm erf}\left(\frac{v_{\rm Gal}}{\sigma_\chi}\right) - \frac{2}{\sqrt{\pi}} \, \frac{v_{\rm Gal}}{\sigma_\chi} \, \exp\left(-\frac{v_{\rm Gal}}{\sigma_\chi}\right)^2\right]~,
    \label{eq:vel_dist_SHM_norm}
\end{equation}
with the same parameters as our simulation and multiplied by the local DM density. We have omitted here the soft-ejection component; as discussed in Sec.~\ref{sec:sim_description}, only the high-energy tail shown above the cutoff is physically well-defined. We have also included the contribution from a \textit{single} black hole bound to Sgr.~A$^\star$ on a circular, 1-yr period orbit. For ease of comparison, we have limited the range of the horizontal axis. However, we emphasize that this last curve extends up to $\sim 8300 \ \rm km/s$ (see App.~\ref{app:M_squared_scaling}). This illustrates how the nuclear star cluster is likely the dominant galactic source of binary-boosted DM, both in terms of flux and ejecta energy. 

For completeness, we note that the escape velocity cutoff in the SHM velocity distribution is understood to result from equilibration during halo formation. The binary boosted component is necessarily unbound and will escape the galaxy, but is quite distinct as it results from an approximately time-independent process after integration over the binary population. Although the impact on the total halo distribution is small, boosting amounts to a reprocessing of a sub-component of the SHM velocity distribution with implications for direct detection experiments.

\section{Extended Direct Detection Sensitivity Prospects}
\label{sec:DD_sensitivity}
Having computed the velocity distribution of the binary-boosted DM component, we now turn to calculate the resulting sensitivity extension of direct detection experiments. We will exclusively focus here on large-volume DM experiments searching for nuclear recoil signals; other possible searches are discussed below. We will begin by outlining the general formalism used to compute recoil rates, and then apply it to a few benchmark classes of DM-SM interactions.

The velocity distributions outlined previously were defined in the halo frame. To compute recoil rates, these must be converted to the laboratory frame, through the transformation $v_\chi \rightarrow |\mathbf{v}_\chi+\mathbf{V}_{\rm det}|$, where $\mathbf{V}_{\rm det}$ is the velocity of the detector relative to the halo. This is typically parameterized as $\mathbf{V}_{\rm det} = \mathbf{V}_{\rm \odot} + \mathbf{V}_{\rm \oplus}(t)$, where there is a fixed contribution from the Sun's motion around the Milky Way, and a periodic contribution from the Earth's motion around the Sun. We neglect the latter, as the amplitude of $|\mathbf{V}_{\oplus}(t)| \sim 30 \ \rm km/s$ is small compared to $|\mathbf{V}_{\odot}| \simeq 232 \ \rm km/s$ (and only relevant for annual modulation effects). Below, we find that this approximation already reproduces the existing direct detection constraints from the SHM with good precision. From here on, we denote by $\mathbf{v}_\chi$ the DM's velocity in the detector's frame.

For a specified differential cross-section $d\sigma/dE_R$, the resulting differential recoil rate is 
\begin{equation}
    \frac{dR}{dE_R} = N_T \, n_\chi \int \eta(E_R) \,  \frac{d\sigma}{dE_R} \, v_\chi \, f(\mathbf{v}_\chi) \, d^3 \mathbf{v}_\chi~,
    \label{eq:diff_recoil_rate}
\end{equation}
where $N_T$ is the total number of targets in the detector, and $f(\mathbf{v}_\chi)$ is the velocity distribution now converted to the detector frame. The function $\eta(E_R)$ is the detector-specific efficiency for observing a nuclear recoil of energy $E_R$. The velocity integral is performed at fixed recoil energy $E_R$, with the implicit lower limit $v_\chi^{\rm (min)}(E_R)$ set by the scattering kinematics. This is specified for the elastic and inelastic cases below. We also note that the integration procedure differs slightly depending on the distribution. In the halo frame, the distribution is isotropic for the SHM and for local binaries. For these cases, upon boosting to the detector frame, the integration is performed over all relative angles between the Earth and DM velocities that kinematically allow for scattering at recoil energy $E_R$. For bulge binaries, however, there is the directionality of the source to consider. In this case, we assume for simplicity that the Sun is on a circular orbit around the Galactic Center, and therefore the velocity of the incoming DM is always orthogonal to the Earth's motion. In practice, this suppresses the differential rate (and the resulting sensitivity) by an approximate factor $\sim 4 \pi$, relative to the isotropic case.

The integrated recoil rate combined with the detector's exposure yields the expected number of nuclear recoil events produced by DM, 
\begin{equation}
    N_\chi = \mathcal{T} \times \int \frac{dR}{dE_R} \,  dE_R~,
    \label{eq:N_chi_detector}
\end{equation}
where $\mathcal{T}$ is the exposure time. The limits of integration above are determined by the support of the efficiency function $\eta(E_R)$, which defines the recoil energy window of the search. We use Eq.~\eqref{eq:N_chi_detector} both to reconstruct existing direct detection constraints assuming the SHM and to compute the extended sensitivity arising from binary-boosted DM. To this end, we evaluate Eq.~\eqref{eq:diff_recoil_rate} replacing either $f(\mathbf{v}_\chi) = f_{\rm SHM}(\mathbf{v}_\chi)$ or $f(\mathbf{v}_\chi) = f_{\rm BBH}(\mathbf{v}_\chi)$.

We specifically consider recent low-mass searches performed by LZ \cite{LZ:2025igz} and PandaX-4T \cite{PandaX:2025rrz}, which focused on the low-energy ionization response of liquid xenon, and achieved detection thresholds as low as $\sim \rm 0.5 \ keV$. Using the reported efficiencies for each search, we compute their sensitivity assuming both the SHM and the various binary-boosted distributions. DarkSide-50 performed a more complex analysis based on the low-energy response of argon and the associated ionization quenching fluctuations. Because of this, we have left an analogous study of DarkSide-50's sensitivity for future work. However, we emphasize that, in principle, this experiment could attain a larger mass reach than either LZ or PandaX-4T for spin-independent scattering.

For LZ, we use the reported upper bound on the number of DM events, and impose this limit on Eq.~\eqref{eq:N_chi_detector} to solve for the cross-section sensitivity. For PandaX-4T, instead we draw the $1\sigma$ confidence level limit from the condition \cite{Raj:2024guv}
\begin{equation}
    \int_0^\infty\left(\frac{\Gamma(N_{\rm obs}+1, N_{\rm 90} + B)}{N_{\rm obs}!}\right) \, \mathcal{N}_{\rm bkg} \, \exp\left(-\frac{\left(B- N_{\rm bkg}\right)^2}{2 \sigma^2_{\rm bkg}}\right) \, dB = 0.1~,
\end{equation}
where
\begin{equation}
    \mathcal{N}_{\rm bkg} = \left[\frac{\sqrt{\pi}}{2} \sigma_{\rm bkg} \left(1 + {\rm erf} \left(\frac{N_{\rm bkg}}{\sqrt{2} \, \sigma_{\rm bkg}}\right)\right)\right]^{-1}~,
\end{equation}
and we use $N_{\rm obs} \simeq N_{\rm bkg} = 322$ as the number of observed/background events. Assuming $\sigma_{\rm bkg} \simeq 0.1$, this yields $N_{\rm 90} \simeq 53$ as the number of DM events we use to set the $1\sigma$ sensitivity line when compared against Eq.~\eqref{eq:diff_recoil_rate}. In each case, using the SHM we reproduce the reported limit from each search within a factor $\sim 2 - 3$, which we find appropriate for the scope of this work. We then repeat this procedure separately using the binary-boosted velocity distributions. For consistency, we do not combine the SHM and boosted components in Eq.~\eqref{eq:diff_recoil_rate}, which  effectively produces a piecewise sensitivity curve. In practice, since the low-mass reach in the SHM is approximately determined by the maximum DM speed in the laboratory frame, combining both distributions yields a similar result.

\subsection{Elastic Spin-Independent Scattering}
We first consider elastic spin-independent interactions, for which we use the standard form for the differential cross-section
\begin{equation}
    \frac{d\sigma_{\chi N}^{\rm (SI)}}{dE_R} = \frac{\sigma^{\rm (SI)}_{\chi n}}{\mu^2_{\chi n}} \frac{m_N}{2 v_\chi^2} A^2 \, F_{\rm SI}^2(q) ~.
    \label{eq:SI-cross-section}
\end{equation}
Above, we fix $A \simeq 131$ as the xenon mass number, $m_N$ is the nuclear mass, $\mu_{\chi n}$ is the reduced DM-nucleon mass, and $\sigma^{\rm (SI)}_{\chi n}$ is the spin-independent reference cross-section at the nucleon level. The Helm form factor $F_{\rm SI}^2(q)$ incorporates the loss of scattering coherence at large momentum transfer, and is parameterized by \cite{Duda:2006uk}
\begin{equation}
    F_{\rm SI}^2(q) = \frac{3j_1(q \, r_N)}{q \, r_N} \ \exp\left[-\frac{(q \, s_N)^2}{2}\right]~,
\end{equation}
where $j_1(x)$ is the spherical Bessel function of the first kind, $r_N \simeq (1.14 \ {\rm fm}) \, A^{1/3}$ is the nuclear radius, $s_N \simeq 0.9 \ \rm fm$, and the momentum transfer magnitude is $q = 2 m_N E_R$. For elastic scattering, we integrate Eq.~\eqref{eq:diff_recoil_rate} from a minimum velocity 
\begin{equation}
    v_{\chi}^{\rm (min)}(E_R) = \sqrt{\frac{m_N E_R}{2 \mu^2_{\chi N}}}~,
\end{equation}
where $\mu_{\chi N}$ is the DM-nucleus reduced mass. 

\begin{figure*}[t!]
    \centering
    \includegraphics[width=0.9\textwidth]{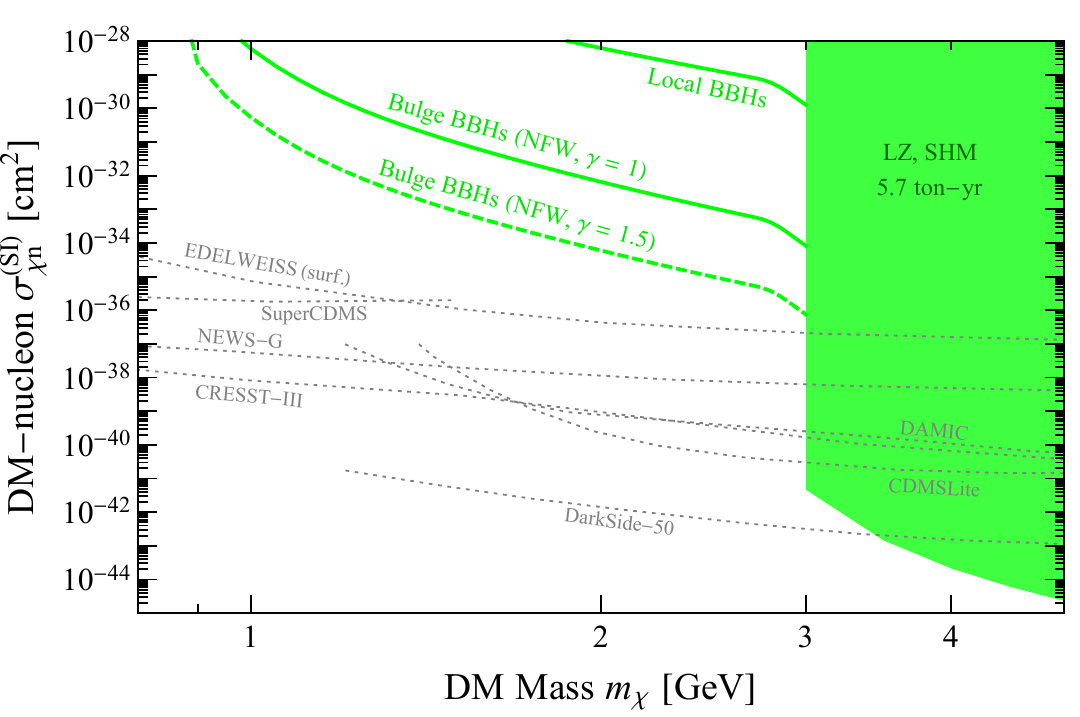}
    \caption{Extended mass sensitivity for spin-independent DM-nucleon scattering for LZ \cite{LZ:2025igz}, from the various black hole binary populations considered here as indicated. The shaded region shows the current limit under the Standard Halo Model. Additional constraints from NEWS-G \cite{NEWS-G:2017pxg}, CRESST-III \cite{CRESST:2019jnq}, DarkSide-50 \cite{DarkSide-50:2022qzh}, DAMIC \cite{DAMIC:2020cut}, CDMSLite \cite{SuperCDMS:2017nns}, and above-ground searches from EDELWEISS \cite{EDELWEISS:2019vjv} and SuperCDMS \cite{SuperCDMS:2020aus} are shown for comparison.}
    \label{fig:LZ_SI_elastic}
\end{figure*}

\begin{figure*}[t!]
    \centering
    \includegraphics[width=0.9\textwidth]{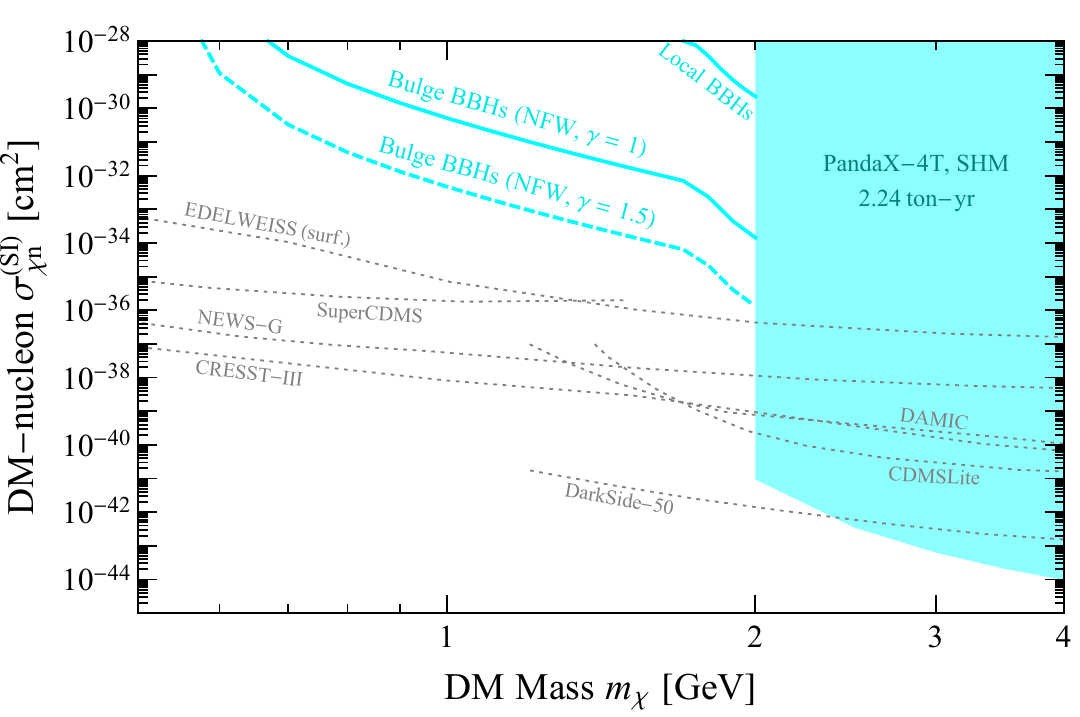}
    \caption{DM-nucleon scattering as in Fig.~\ref{fig:LZ_SI_elastic}, but now showing the extended mass sensitivity of PandaX-4T \cite{PandaX:2025rrz}.}
    \label{fig:PandaX_SI_elastic}
\end{figure*}

Figures~\ref{fig:LZ_SI_elastic} and \ref{fig:PandaX_SI_elastic}, respectively, show the extended sensitivity of LZ and PandaX-4T to the flux of gravitationally-boosted DM from binary black holes, along with reported limits from various collaborations operating low-threshold experiments. For the distributions derived from bulge binaries, it can be seen that the DM mass reach of LZ and PandaX-4T can be extended down to $\sim 0.9 \rm \ GeV$ and $\sim 0.6 \rm \ GeV$. While this reach technically goes further, we terminate the lines at a cross-section $\sim 10^{-28} \ \rm cm^2$. This serves as a convenient approximate threshold above which energy loss in the rock overburden cannot be neglected. While lower-threshold experiments remain generally more sensitive, we emphasize that our results are based on a limited subset of the Galactic black hole binaries, and do not include the potentially dominant contribution from the Milky Way’s nuclear star cluster. Including this source, in particular, would further enhance the boosted flux and could allow experiments such as PandaX-4T (and likely DarkSide-50) to probe masses beyond the lowest values that currently \textit{only} CRESST-III is able to access via nuclear scattering. More broadly, these results demonstrate that gravitational boosting by black hole binaries can provide a competitive and largely particle-physics-independent mechanism for extending the mass reach of direct detection experiments.

\subsection{Elastic Spin-Dependent Scattering}
We next consider the case of DM scattering through spin-dependent interactions. We will exclusively focus on PandaX-4T here \cite{PandaX:2025rrz}, which achieved a similar spin-dependent cross-section sensitivity to LZ while probing DM masses down to $2 \ \rm GeV$. We do not expect argon-based DarkSide-50 to be sensitive in this case, since argon nuclei carry no net spin (this detector could still probe spin-dependent scattering via loop-induced operators \cite{Bozorgnia:2024kkf}, although we do not consider this possibility here). 

In this scenario, we now parameterize the differential cross-section as
\begin{equation}
    \frac{d\sigma_{\chi N}^{\rm (SD)}}{dE_R} = \frac{\sigma^{\rm (SD)}_{\chi (p,n)}}{\mu^2_{\chi n}} \frac{m_N}{2 v_\chi^2} \frac{4 \pi }{3 (2 J + 1)}\, S_{\rm (p,n)}(q)~,
    \label{eq:SD-cross-section}
\end{equation}
where $J$ is the target's nuclear spin. As in the spin-independent case, $m_N$ is the nuclear mass, $\mu_{\chi n}$ is the DM-nucleon reduced mass, and $v_\chi$ is the DM velocity in the detector's frame. We consider two benchmark scenarios in which the interaction is dominated by either the proton or neutron spin contribution, denoted by $(p,n)$; note that this choice does not imply that the coupling is exclusively to protons or neutrons.
The function $S_{(p,n)}(q)$ is the corresponding momentum transfer–dependent nuclear structure function, and $\sigma^{\rm (SD)}_{\chi (p,n)}$ is the reference nucleon-level cross section at zero momentum transfer. For the former, we use the calculations from Ref.~\cite{Klos:2013rwa}. The two relevant isotopes are $^{129}$Xe and $^{131}$Xe, which respectively have $J = 1/2$ and $J = 3/2$, and constitute approximately half of the total detector's mass. Because we are considering elastic scattering, the same kinematic limits apply as in the previous case. 

\begin{figure*}[t!]
    \centering
    \includegraphics[width=0.9\textwidth]{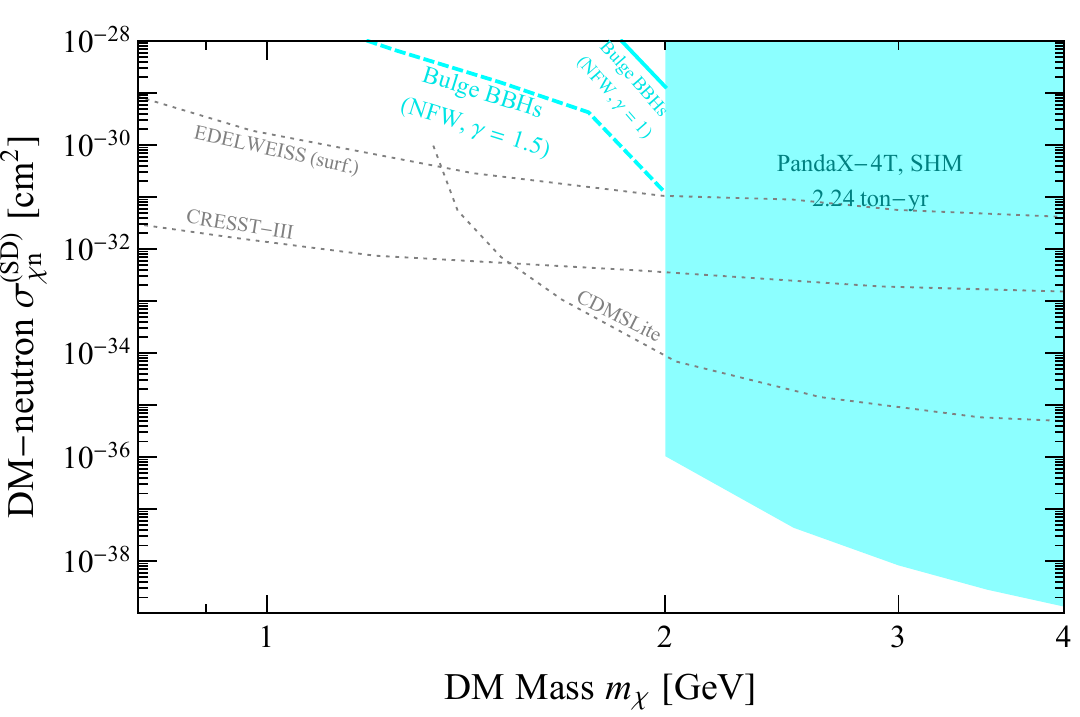}
    \caption{Extended mass sensitivity for spin-dependent neutron-only scattering for PandaX-4T \cite{PandaX:2025rrz}, from the various binary populations considered here as indicated. The shaded region shows the current limit under the Standard Halo Model. Additional limits assuming the SHM from CRESST-III \cite{CRESST:2019jnq}, CDMSLite \cite{SuperCDMS:2017nns}, and an above-ground search from EDELWEISS \cite{EDELWEISS:2019vjv} are shown.}
    \label{fig:SD_neutron}
\end{figure*}

Figure \ref{fig:SD_neutron} shows the extended mass sensitivity of the PandaX-4T experiment to the benchmark neutron-only spin-dependent interactions, as a function of DM mass. Depending on the assumed DM halo profile, the binary-boosted flux can extend the accessible mass of this experiment down to $\sim 1.2 \ \rm GeV$ in the neutron-only case. We have not included the contribution from local black hole binaries, as this is sufficiently weak that it requires cross-sections above overburden limits to be detectable in this case. For comparison, additional SHM-only limits from various experiments with suitable scattering targets are shown in Figure \ref{fig:SD_neutron}. While these experiments typically probe smaller cross sections, this comparison highlights that gravitational boosting by black hole binaries can render a large-scale detector such as PandaX-4T nearly competitive with dedicated low-threshold systems in a largely model-independent manner. In principle, the same analysis can be applied for these low-threshold experiments. However, even for the most optimistic flux estimate we have obtained here, the extended sensitivity would lie at cross-sections larger than $\sim 10^{-28} \ \rm cm^2$. We have also explored the corresponding sensitivity to proton-only spin-dependent interactions, but do not find an enhancement to mass sensitivity for cross-sections below $\sim 10^{-28} \ \rm cm^2$ and thus do not show this case explicitly. For completeness, we mention that in the case of proton-only scattering, NEWS-G \cite{NEWS-G:2024jms} places the dominant bound in the mass range shown in Fig.~\ref{fig:SD_neutron}, although we reach the same conclusion in terms of its potential reach upon performing a similar analysis. As before, incorporating the full contribution from the Milky Way’s nuclear star cluster could further enhance the boosted flux, potentially extending the mass reach well into the sub-GeV regime.

\subsection{Inelastic Scattering}
Lastly, we consider inelastic scattering, whereby a transition is induced between DM states of different mass, $i.e.$ $\chi_{1(2)} + N \rightarrow \chi_{2(1)} + N$, with a mass splitting 
\begin{equation}
    \delta = m_{\chi_2} - m_{\chi_1}~.
\end{equation}
For this scenario, we adopt the same spin-independent cross-section parameterization of Eq.~\eqref{eq:SI-cross-section}. The kinematics, however, will now qualitatively differ from the elastic case, as a minimum incoming DM velocity is required to excite the heavier state. 

In the laboratory frame, the maximum mass splitting that can be excited for non-relativistic DM is 
\begin{equation}
    \delta_{\rm max} = \frac{1}{2} \, \mu_{\chi N} \, v_\chi^2 \simeq 384 \ {\rm keV} \, \left(\frac{\mu_{\chi N}}{122.6 \ \rm GeV}\right) \left(\frac{v_\chi}{750 \ \rm km/s}\right)^2~,
    \label{eq:deltamchi_max}
\end{equation}
where, on the right hand side, we have normalized the reduced DM-nucleus mass to that of a xenon nucleus (corresponding to the limit $m_\chi \gg m_N$), and $v_\chi$ to approximately the maximum speed the DM can have under the SHM in the detector's frame. This simple kinematic threshold has important implications for direct detection: given the typical speed of halo DM, endothermic scattering becomes kinematically forbidden in xenon-based detectors for mass splittings above a few hundred keV, since the collision energy is insufficient to up-scatter the DM into the heavier state. Experiments employing lighter target nuclei, like argon-based DarkSide-50, are subject to an even more restrictive mass splitting bound due to the smaller reduced mass, and therefore we will not consider them in this section. In models where the relic abundance of the heavier state $\chi_2$ is depleted through decays $\chi_2 \rightarrow \chi_1 + (\rm additional \ states)$, exothermic scattering, which does not require an energy threshold, cannot occur either since the initial state $\chi_2$ is absent. As a consequence, if the mass splitting is large enough, standard nuclear recoil searches lose significant sensitivity to this class of models, as the only remaining kinematically allowed interaction is loop-induced elastic scattering and therefore suppressed. Inelastic DM scenarios have been widely studied as a means of reconciling null direct detection searches with astrophysical excesses potentially associated with DM, see $e.g.$ \cite{Tucker-Smith:2001myb,Chang:2008gd,Finkbeiner:2007kk,Pospelov:2007xh,Batell:2009vb,Ghorbani:2014gka,Zhang:2016dck,Alvarez:2019nwt,Hooper:2025fda}.

\begin{figure*}[t!]
    \centering
    \includegraphics[width=0.9\textwidth]{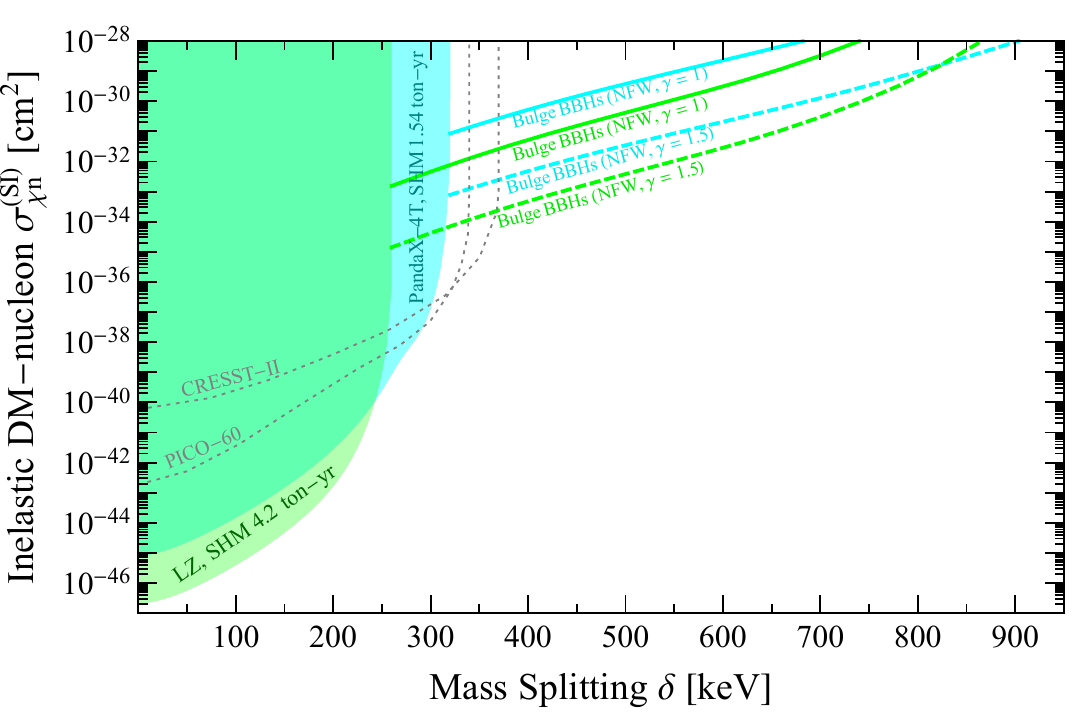}
    \caption{Extended mass splitting sensitivity for spin-independent inelastic scattering, assuming a DM mass $m_\chi = 1 \ \rm TeV$, from the various binary populations considered here as indicated. Shaded regions denote the sensitivity of PandaX-4T and LZ under the Standard Halo Model. Additional limits under the SHM from PICO and CRESST-II are shown \cite{Bramante:2016rdh}.}
    \label{fig:SI_inelastic}
\end{figure*}

To compute the differential recoil rate, we use Eq.~\eqref{eq:N_chi_detector} combined with the differential cross-section Eq.~\eqref{eq:SI-cross-section} as in the spin-independent elastic case (for simplicity, we do not consider here spin-dependent inelastic interactions, although the analysis for applicable models would be similar). However, Eq.~\eqref{eq:diff_recoil_rate} is now integrated from a minimum velocity that, in the limit $m_N \ll m_\chi$, reads
\begin{equation}
    v^{\rm (min)}_{\chi}(E_R) \simeq \frac{1}{\sqrt{2 m_N E_R}} \left(\frac{E_R \, m_N}{\mu_{\chi N}}+\delta\right)~,
\end{equation}
which is the minimum velocity required to produce a recoil energy $E_R$ given a finite mass splitting $\delta$. 

By incorporating the flux of binary-boosted particles, the maximum mass splitting that direct detection experiments can probe is substantially increased, as the enhanced velocities translate directly into larger available energy for inelastic scattering. Importantly, unlike other proposed DM-boosting mechanisms where the energy gain rapidly drops with increasing DM mass, the binary-boost remains efficient even for heavy DM, enabling direct detection experiments to probe inelastic DM models with large mass splittings \textit{at high masses}. We therefore restrict our attention to this parameter space in what follows. 

Figure~\ref{fig:SI_inelastic} shows the extended sensitivity to inelastic DM splitting of PandaX-4T and LZ, for a fixed TeV DM mass. For comparison, we have also included the corresponding SHM–only sensitivities, derived from published results of both collaborations with exposures of $1.54$~ton-year for PandaX-4T \cite{PandaX:2024qfu} and $4.2$~ton-year for LZ \cite{LZ:2024zvo}. Assuming only the SHM, the maximum mass splitting probed is set by the recoil energy search window. PandaX-4T performed a relatively wider search up to $\sim 120 \ \rm keV$ recoils compared to LZ's $\sim 70 \ \rm keV$, therefore achieving slightly larger coverage. We have also plotted derived limits from CRESST-II and PICO-60 under the SHM assumption, which probe even higher mass splittings, but are otherwise unsuitable for this analysis due to their smaller cross-section sensitivity. For CRESST-II, the larger splitting reach is because of their use of tungsten as a scattering target, whereas PICO-60 used a much wider search window extending to $\rm MeV$-scale nuclear recoils. Gravitationally-boosted DM, which acquires much larger velocities than its halo counterpart, significantly extends the kinematically-accessible mass splittings, potentially probing up to $\sim 800 - 900 \ \rm keV$ for bulge binaries. While the mass-splitting reach extends further, we have terminated the plots also at $10^{-28} \ \rm cm^2$ as in the prior cases. Moreover, in this heavy DM mass regime, there exists an upper limit on the cross section beyond which the mass number scaling assumed in Eq.~\eqref{eq:SI-cross-section} is no longer generic \cite{Digman:2019wdm}, although explicit models that preserve this functional dependence at larger cross sections can be constructed \cite{Acevedo:2024lyr}.
We reiterate that, as in the elastic cases, we have not considered the contribution from the nuclear star cluster, which possibly extends the reach well into the MeV-splitting scale, given that we have numerically observed ejections with velocities of up to $\sim 8300 \ \rm km/s$ in this case (see App.~\ref{app:M_squared_scaling}).

Finally, we note that there exist additional putative limits on inelastic DM scattering from astrophysical probes. These include white dwarf heating \cite{McCullough:2010ai}, the orbital evolution of the Galactic Center gas cloud G2 \cite{Acevedo:2025rqu}, annihilation around neutron stars \cite{Acevedo:2024ttq}, kinetic heating of neutron stars \cite{Baryakhtar:2017dbj,Acevedo:2019agu,Alvarez:2023fjj}, and the cooling of active galactic nuclei \cite{Gustafson:2025dff}. While some of these searches can have larger reach in mass splitting, they can also be less robust in terms of the assumed DM density around the probe and, in some cases, yield weaker cross-section constraints. A recent analysis \cite{Alam:2025qbq} has also extended previous limits from the Xenon-100 experiment based on DM up-scattering by lead within the Earth's crust, combined with the impact of the large Magellanic cloud on the halo distribution \cite{Smith-Orlik:2023kyl,Reynoso-Cordova:2024xqz}, reaching splittings of order $\sim 800 \ \rm keV$.

\subsection{Other Potential Detectors}
Since the gravitational boosting mechanism discussed in this work is largely mass-independent and does not make specific assumptions about the nature of the DM interaction with the SM, it is worth considering other potential detection approaches. 

One possibility is the use of neutrino detectors to search for a binary-boosted DM flux, considering the maximum energy deposited in a nuclear recoil is
\begin{equation}
    E^{\rm max}_{R} = \frac{2 \mu^2_{\chi N} v_\chi^2}{m_N} \simeq \left(\frac{v_\chi}{8300 \ \rm km/s}\right)^2 \times \begin{cases}
      \ 0.11 \ {\rm MeV} \, \left(\frac{m_\chi}{\rm GeV}\right)^2 \left(\frac{m_N}{12 \ \rm GeV}\right)^{-1} & \ m_\chi \ll m_N \\
      \\
      \  17.2 \ {\rm MeV} \, \left(\frac{m_N}{12 \ \rm GeV}\right)  & \ m_\chi \gg m_N
    \end{cases} \ \ \ \ ,
\end{equation}
where we have normalized the DM velocity in the detector's frame to about the maximum we have numerically observed for a black hole orbiting Sgr.~A$^\star$ ($cf.$ App.~\ref{app:M_squared_scaling}), and taken for reference the target mass of a carbon nucleus. The recoil energy in the light DM case appears too low to be directly observed in neutrino detectors, even those with the lowest thresholds such as Borexino or JUNO. However, these experiments could still be sensitive to annual modulation signals rather than observable single-scatter events \cite{Leane:2025efj}.  By contrast, the energy transfer scale of a DM-nucleus collision is in principle large enough for it to be detected in the regime $m_\chi \gg m_N$. We leave for future investigation the question of whether novel sensitivity can be attained with neutrino detectors, pending a full computation of the binary-boosted flux from the Milky Way's nuclear star cluster.

Another possibility is mineral-based detectors, in which certain naturally occurring materials excavated from the Earth's crust preserve damage tracks created by transiting high energy particles over geological timescales \cite{Drukier:2018pdy,Baum:2018tfw,Edwards:2018hcf,Baum:2019fqm,Baum:2022wfc,Ebadi:2021cte,Acevedo:2021tbl,Caccianiga:2024otm}. These mineral samples can be dated to estimate their exposure, and their tracks (along with radiogenic backgrounds) analyzed through multiple microscopy techniques, see $e.g.$ Ref.~\cite{Baum:2023cct}. In this scenario, binary-boosted DM would be able to cause observable damage more easily to the sample, given the enhanced kinetic energies. In the light DM regime, we expect a relatively similar analysis would extend the sensitivity of these searches, which may be competitive with low-threshold direct detection experiments depending on future improvements of the read-out technologies and sample size \cite{Fung:2025cub}. On the other hand, it is possible that the enhanced energies from the binary-boosting process might also improve the sensitivity to heavy DM in the multi-scatter regime, see $e.g.$ Refs.~\cite{Ebadi:2021cte,Acevedo:2021tbl,Bozorgnia:2025lsl}, a region of parameter space where these searches may have advantages as compared to the detection approaches that we have considered here. 

We have also left for future investigation other processes which, combined with this novel boosting channel, may further increase the sensitivity of available experiments. This includes, for example, the Migdal effect \cite{Ibe:2017yqa,Dolan:2017xbu,Baxter:2019pnz,Bell:2019egg, Essig:2019xkx,GrillidiCortona:2020owp,Bell:2021zkr,Acevedo:2021kly}, as well as bremsstrahlung \cite{Kouvaris:2016afs, Bell:2019egg, GrillidiCortona:2020owp}, both of which have an inherent velocity dependence in the scattering rate that could potentially be enhanced with boosted DM. 

Finally, we comment on DM-electron scattering as an alternative interaction channel. This process is considerably more constraining compared to DM-nucleus scattering for sub-GeV masses, partly due to the closer mass matching with the electron which leads to more favorable scattering kinematics for detection. Although a similar analysis of binary boosted MeV DM could in principle extend direct detection limits down to the sub-MeV regime, this mass range and the cross-section sensitivity that is currently achievable with the boosted flux is subject to additional strong limits from stellar and supernovae energy loss \cite{Dent:2012mx,An:2013yfc,Hardy:2016kme}, and cosmological observables \cite{Viel:2013fqw, Sabti:2019mhn, Buen-Abad:2021mvc}. While some of these constraints can be model-dependent, they restrict the model space probed by the binary-boosted flux. For this reason, we have not considered DM-electron scattering in detail in this work. 

\section{Conclusions and Outlook}
\label{sec:outlook}
In this paper, we have investigated the gravitational interactions between DM and binary systems in the Milky Way, and identified a new mechanism for accelerating DM particles. Utilizing simple Monte Carlo simulations, combined with analytic arguments to interpret the outputs, we have shown how compact black hole binaries are especially effective at transferring energy to transiting DM particles, through close gravitational deflections deep within the binary's potential. Because of its gravitational nature, this mechanism is independent of the assumed DM-SM interaction portal, and largely insensitive to the DM mass. The only assumptions we have made are negligible DM self-interactions, and a mass range that avoids both the wave-like regime and significant back-reaction on the binary. 

We have numerically computed the ejection spectra for a range of black hole binary periods, and their associated velocity distributions. We have combined these with existing synthetic catalogues to estimate the collective impact from these systems across the Galaxy on the DM velocity distribution. However, we have not attempted to be exhaustive in terms of the full binary population. For simplicity, we focused on a few representative examples: the expected local black hole binaries within $\sim 1\ \rm kpc$ from Earth; binaries residing in the bulge within $\sim 1\ \rm kpc$ from the Galactic Center; and a \textit{single} black hole bound to Sgr.~A$^\star$ on a 1-yr orbit. The resulting distribution on Earth, while comparatively small in terms of flux, achieves velocities up to $\sim 2000 \ \rm km/s$ for double black hole binaries, and $\sim 8300 \ \rm km/s$ for black holes bound to Sgr.~A$^\star$. The last case, in particular, points to the Milky Way's nuclear star cluster as a potentially dominant source of binary-boosted DM, both in terms of flux and ejection energy. This follows from the unique combination of the highest DM densities in the Galaxy, a large population of stellar-mass black holes, and the deep gravitational potential of Sgr.~A$^\star$, which substantially enhances the maximum ejection velocities. A dedicated analysis of the black hole population in this region is left for future investigation \cite{Acevedo:2026tur}.

The binary-boosted flux can extend the mass sensitivity of several direct detection experiments in a largely model-independent way. As a first application of this mechanism, we have focused on DM-nucleus scattering inside the large-volume detectors LZ and PandaX-4T. Utilizing existing ionization searches from these collaborations, we have shown how binary-boosted DM enhances their sensitivity down to the sub-GeV regime for spin-independent and spin-dependent scattering. In addition, given that this boosting mechanism is practically independent of DM mass, we have shown how it can substantially expand the kinematic space for inelastic DM searches at heavy DM masses. This approach can be extended to a number of other detectors that may allow the mass reach to be extended further.  In particular, we have left DarkSide-50 for a future analysis, as it requires a more careful computation of the argon's response at low energy transfers, but we highlight that this experiment may exhibit a higher sensitivity in both cross-section and mass in the sub-GeV regime compared to PandaX-4T or LZ. Additional lower-threshold experiments, such as CRESST and Super-CDMS, can in principle probe even lower masses through this DM-boosting channel. However, given their smaller fiducial volumes, our conservative estimates suggest that the required cross-sections would likely be too large to ignore energy loss through the overburden. A full exploration of the sensitivity of these lower-threshold detectors, as well as alternative targets such as neutrino observatories or mineral-based experiments, including contributions from the Milky Way’s nuclear star cluster, is left for future work.

Beyond terrestrial DM detectors, several other processes merit investigation. These include the reverse process of gravitational capture and subsequent annihilation of DM, potential enhancements to dark kinetic heating in compact binary systems, and the back-reaction of DM on binary dynamics. Investigating these avenues will provide a more complete understanding of the astrophysical consequences of DM interactions with binaries, and may reveal additional indirect signatures.

\section*{Acknowledgements}
We thank Aleksandra Olejak for helpful discussions about synthetic binary black hole catalogues. This research was undertaken thanks in part to funding from the Natural Sciences and Engineering Research Council of Canada through the Arthur B. McDonald Canadian Astroparticle Physics Research Institute. This research was also enabled in part by support provided by Compute Ontario (computeontario.ca) and the Digital Research Alliance of Canada (alliancecan.ca). Simulations were performed on the Alliance's \textit{Nibi} and \textit{Trillium} computing clusters. 

\appendix

\bigskip

\noindent {\bf{\large Appendices}}

\section{Sample Particle Trajectories}
\label{app:DM_orbits}
\begin{figure*}[t!]
    \centering

    \begin{subfigure}{0.49\textwidth}
        \centering
        \includegraphics[width=\textwidth]{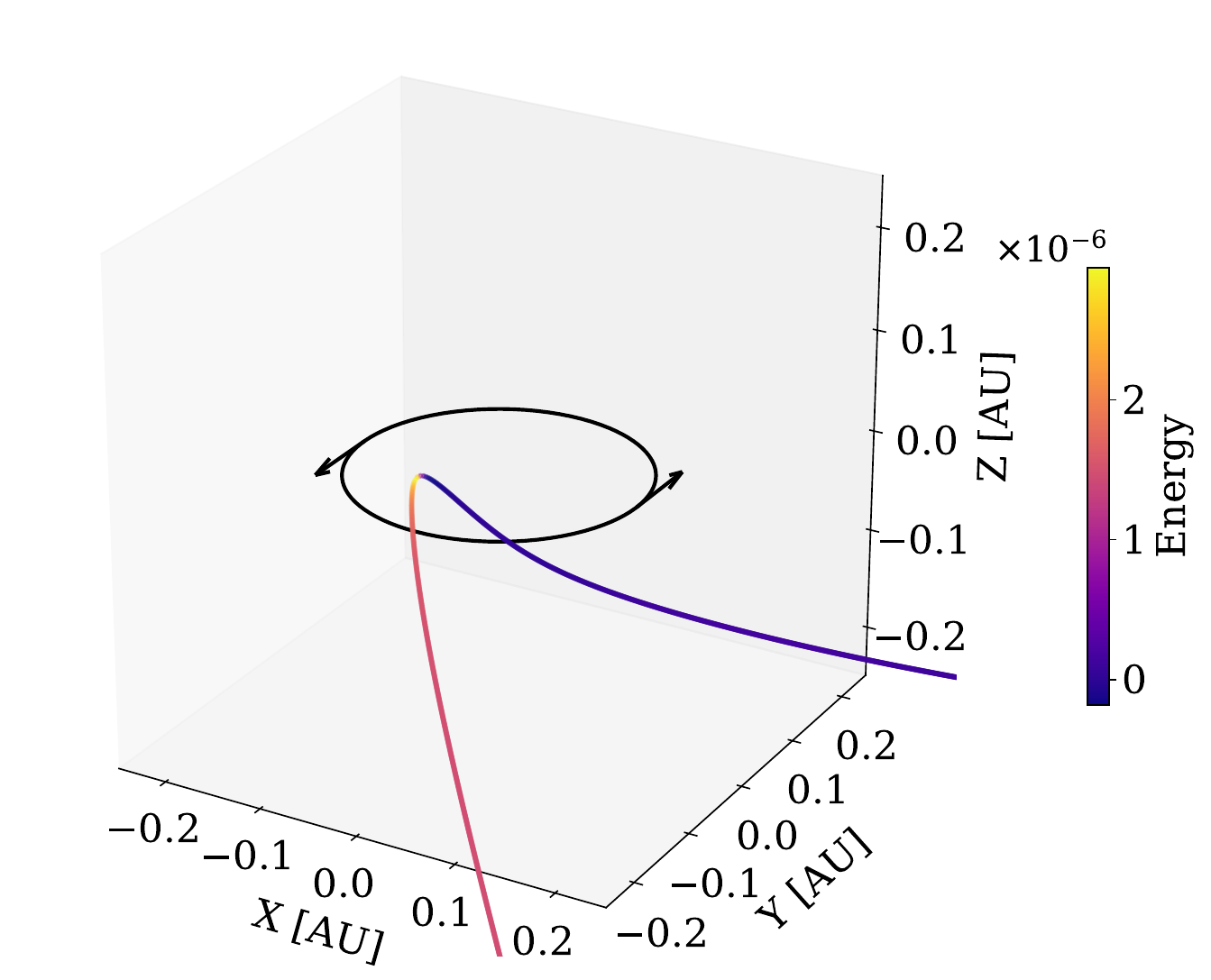}
    \end{subfigure}
    \hfill
    \begin{subfigure}{0.49\textwidth}
        \centering
        \includegraphics[width=\textwidth]{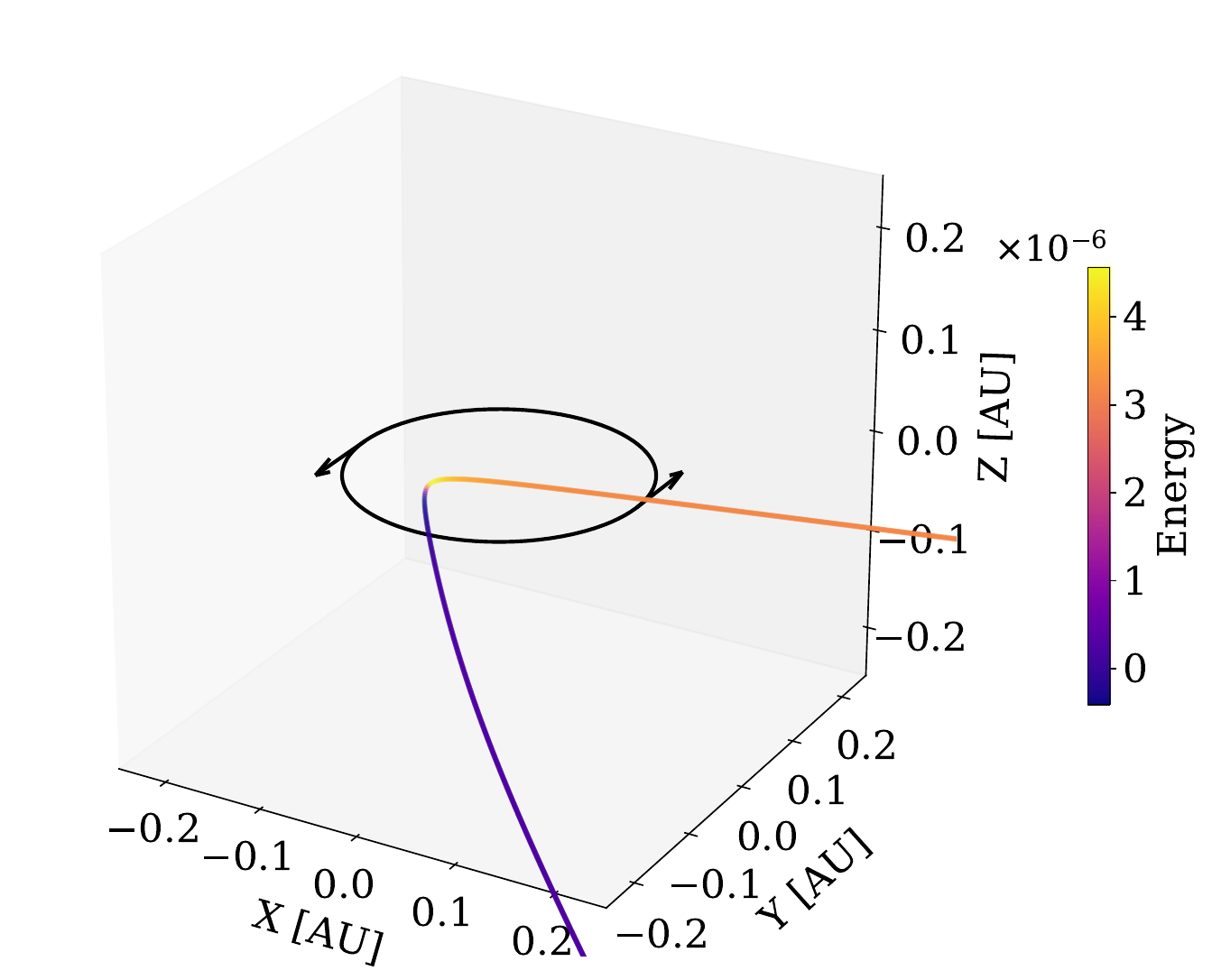}
    \end{subfigure}

    \vspace{0.2cm}

    \begin{subfigure}{0.49\textwidth}
        \centering
        \includegraphics[width=\textwidth]{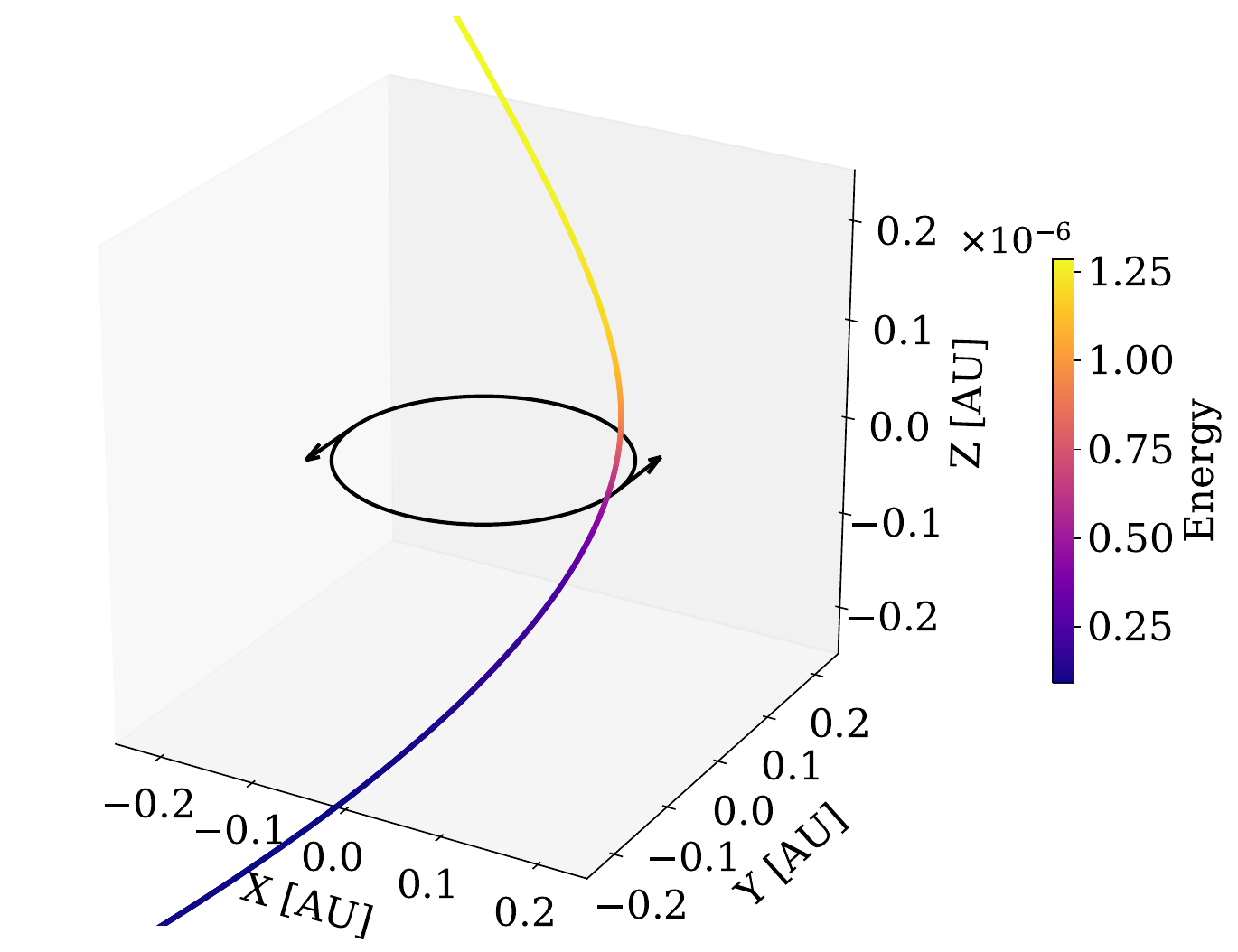}
    \end{subfigure}
    \hfill
    \begin{subfigure}{0.49\textwidth}
        \centering
        \includegraphics[width=\textwidth]{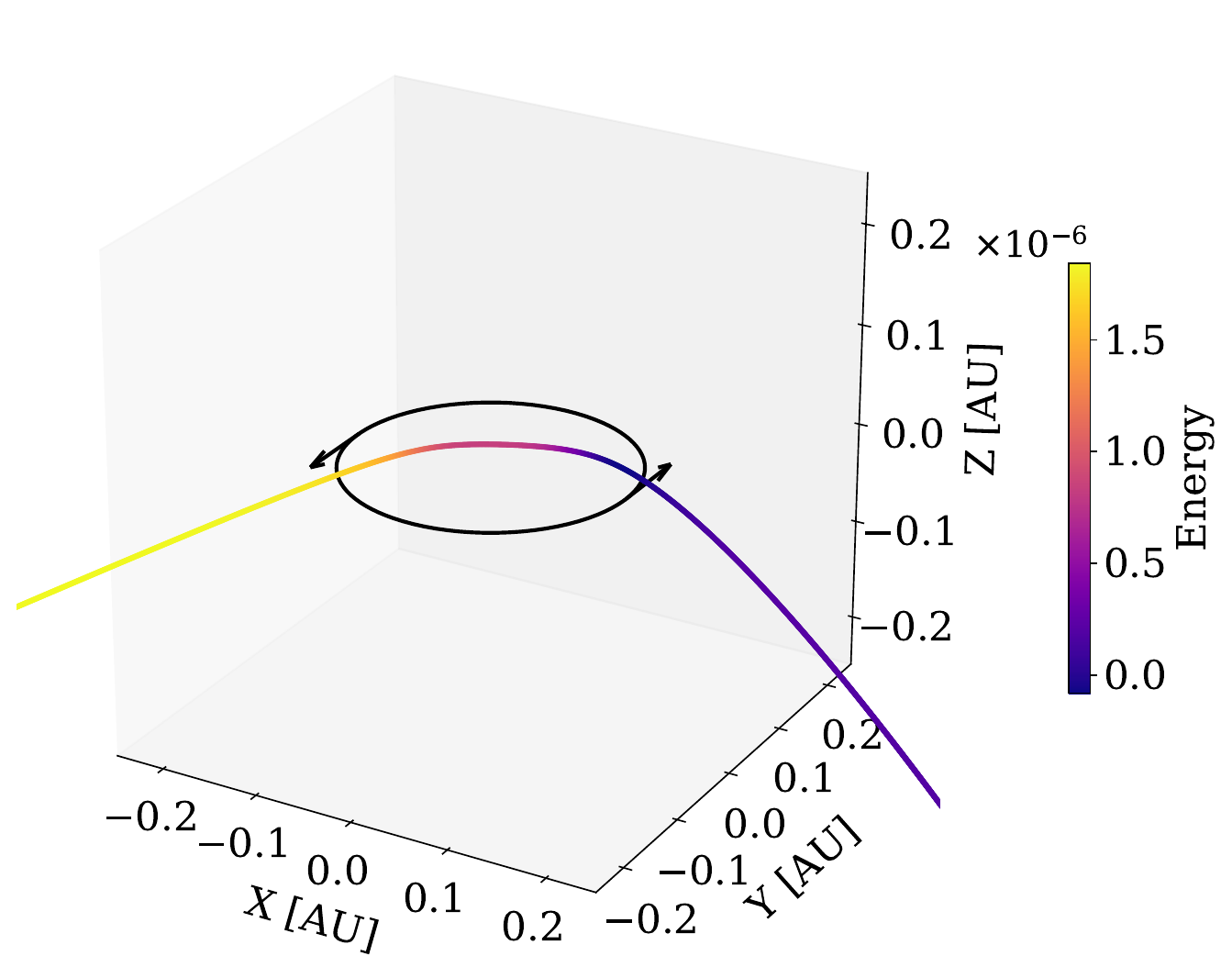}
    \end{subfigure}

    \vspace{0.4cm}

    \caption{Sample of simulated trajectories with net energy gain, for a $P = 10^{-2} \ \rm yr$, equal-mass, circular black hole binary with $M_{1} + M_{2} = 28 \, M_\odot$. The trajectories are colored according to the instantaneous energy of the particle at each point. The black ring in the $Z = 0$ plane marks the trajectory traced by the binary components, with the arrows indicating the direction of orbital motion.}

    \label{fig:example_trajectories}
\end{figure*}

Figure~\ref{fig:example_trajectories} shows a sample of simulated trajectories for the DM particles, for a fiducial equal-mass, circular, black hole binary ($cf.$ Fig.~\ref{fig:BBH_Spectra}). The binary's orbit is indicated by the black central ring in the $Z = 0$ plane, with the arrows showing the spin direction. In all cases, it can be seen that gravitational scattering that deflects the incoming particle (at least partially) in the direction of motion of one of the binary components results in energy gain.   

\section{Energy and Angular Distribution Approximations}
\label{app:ejecta_dist_approx}
\subsection{Energy/Angle Factorization}
\begin{figure*}[t!]
    \centering
    \hspace*{-2.0cm}
    \includegraphics[width=1.25\textwidth]{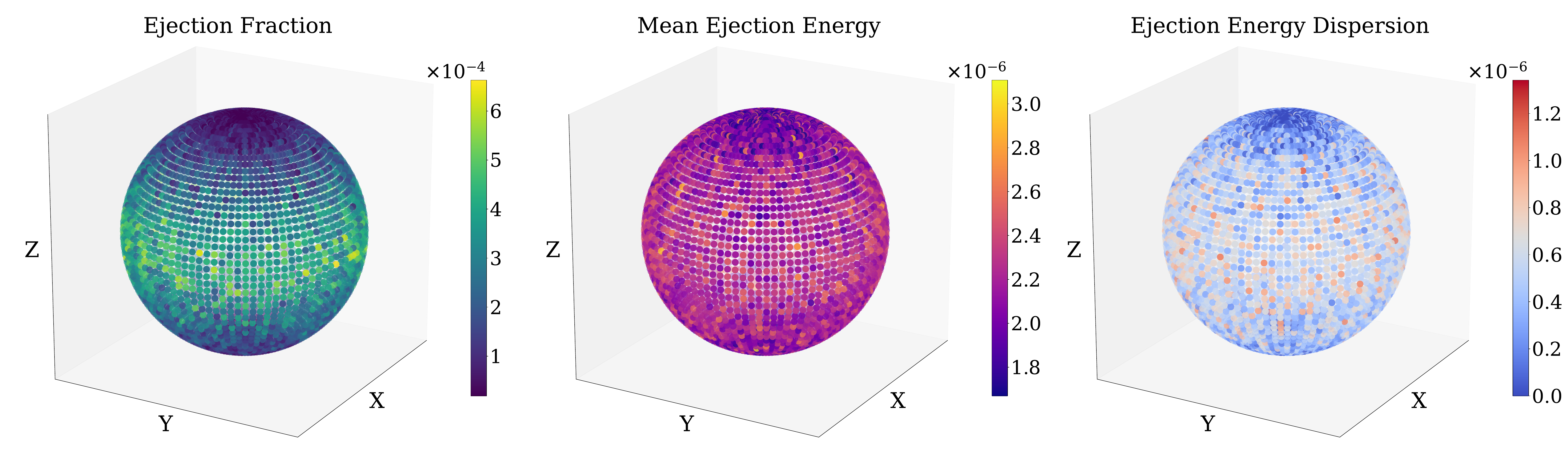}
    \caption{Angular distribution of the number of ejected particles (\textit{left}), mean ejection energy (\textit{middle}) and the dispersion in ejection energies (\textit{right}), for a $P = 10^{-2} \, \rm yr$, equal-mass, circular black hole binary with $M_{1} + M_2 = 28 \, M_\odot$ assuming its barycenter is at rest relative to the halo. The binary plane corresponds to the equator of the unit sphere. For reference, each dot is an angular bin in spherical coordinates of size $\Delta\phi = \Delta\theta = \pi/50 \, \, (= 3.6^\circ)$.}
    \label{fig:ejecta_ang_dist_no_Vcm}
\end{figure*}

Figure~\ref{fig:ejecta_ang_dist_no_Vcm} shows the angular distribution of the ejecta in terms of the fraction of recorded ejections, average energy, and energy dispersion, for the same fiducial black hole binary of Fig.~\ref{fig:example_trajectories}. To construct these plots, we restrict the sample to those particles ejected with $\Delta \varepsilon > 0$ such that their final energy is above the escape cutoff $\varepsilon \simeq 1.65 \times 10^{-6}$. This effectively removes a large fraction of the ejecta that underwent weak and distant encounters with the binary. 

It is clear that, in percentage terms, ejections preferentially occur towards the binary's plane at $Z = 0$ (the equator in the unit sphere), despite the particles being initialized with isotropic incident directions at the simulation boundary. To understand this, we first note that the vast majority of deflections observed are weak, corresponding to small scattering angles, and so particles injected perpendicular or parallel to the binary's axis are also the dominant contribution to the ejected particle count along these directions. However, particles initially incident from directions nearly aligned with the orbital plane have a larger probability of being deflected with energy gain, as there can be a considerable overlap with the trajectory swept out by either binary component, thereby increasing the gravitational interaction time. By contrast, particles incident nearly perpendicular to the binary plane have a smaller chance of being deflected with an energy gain, as now the interaction time is generally reduced given that they move mostly perpendicular to the motion of either companion. Formally, the ejection fraction with $\Delta \varepsilon > 0$ on the binary's axis vanishes by definition, since the final velocity has no projection onto either companion's motion, but the simulation records a finite number as a consequence of finite angular binning size (see the next subsection).

In terms of the average ejection energy, we find less preference for a specific direction, and only a modest decrease in ejecta towards the binary axis. The energy dispersion does not exhibit a strong directional preference either; in this case, the lower dispersion towards the poles of the unit sphere is driven by the lower sample of recorded ejecta. Consequently, we opt to factorize the joint probability distribution of angle and energy, Eq.~\eqref{eq:factorize_energy_angle}, as an approximation to the true function, which significantly reduces simulation time given our binning sizes and error tolerance. 

\subsection{Isotropic Ejection}
In Eq.~\eqref{eq:isotropic_prob}, we have further approximated the angular distribution of the ejecta as isotropic. This approximation can be tested against the simulated angular distributions shown in Fig.~\ref{fig:ejecta_ang_dist_no_Vcm}. The numerical distribution is normalized such that
\begin{equation}
  \sum_{i,j} \frac{N(\theta_i,\phi_j)}{N_{\rm ej}} =  \sum_{i,j} \frac{1}{\Delta \Omega_i} \left(\frac{N(\theta_i,\phi_j)}{N_{\rm ej}}\right) \, \Delta\Omega_i = 1 ~,
\end{equation}
whereas in the isotropic case
\begin{equation}
    \int \frac{dp}{d\Omega} \, d\Omega = 1~,
\end{equation}
with $dp/d\Omega = 1/4\pi$. 

Let us consider both the edge-on and face-on directions relative to the binary plane. To construct the angular distribution, we have binned both angular directions in steps of size $\delta \theta = \delta \phi = \pi/50$, so that $\Delta \Omega_i = (\pi/50)^2 \sin\theta_i$. The ejection fraction is of order $\sim 5 \times 10^{-4}$ parallel to the binary plane ($\sin \theta = 1$), implying a numerical probability in this direction of order 
\begin{equation}
    \frac{1}{\Delta\Omega_{\theta = \pi/2}} \frac{N(\theta = \pi/2, \phi)}{N_{\rm ej}} \simeq 0.126~,
    \label{eq:dpdomega_max}
\end{equation}
which is slightly larger than just assuming isotropic emission ($1/4\pi \simeq 0.079$). On the other hand, in the strictly face-on directions ($\sin \theta = 0$), the ejection fraction is formally zero, since the deflections by definition have no projection onto the direction of motion of either companion, and therefore do not physically result in any change in energy, $cf.$ Eq.~\eqref{eq:delta_epsilon}. However, this only applies to a measure-zero set and, in any case, we expect any realistic system to possess a finite inclination. The simulation records a finite number of ejections as a consequence of the finite angular binning, which naturally also includes off-axis directions. Taking at face value the ejection fraction recorded in the first angular bin in $\theta$, of order $\sim 5 \times 10^{-6}$, we obtain
\begin{equation}
    \frac{1}{\Delta\Omega_{\theta \simeq \delta\theta/2}} \frac{N(\theta \simeq \delta\theta/2, \phi)}{N_{\rm ej}} \simeq 0.040~,
    \label{eq:dpdomega_min}
\end{equation}
which is marginally smaller than the isotropic factor $1/4\pi$. Given the relatively small angular binning, we expect most binaries to have angular distributions within the range spanned by Eqs.~\eqref{eq:dpdomega_max} and \eqref{eq:dpdomega_min}. Moreover, upon integrating over a large binary population, we expect this approximation to average out, since it overestimates the flux from binaries with face-on orbits but also underestimates the flux of those binaries with edge-on orbits relative to the line of sight. While we have performed these estimates for a benchmark binary of fixed period, the same qualitative conclusion is reached for all the systems we have simulated in this work. 

\section{Ejection Spectra with Additional Binary Parameters}
We analyze here the properties of the ejection spectra accounting for additional binary parameters that we suppressed in our main analysis: eccentricity, mass ratio, and barycenter motion across the halo. 

\subsection{Eccentricity}
\label{app:eccec}
\begin{figure*}[t!]
    \centering
    \includegraphics[width=0.9\textwidth]{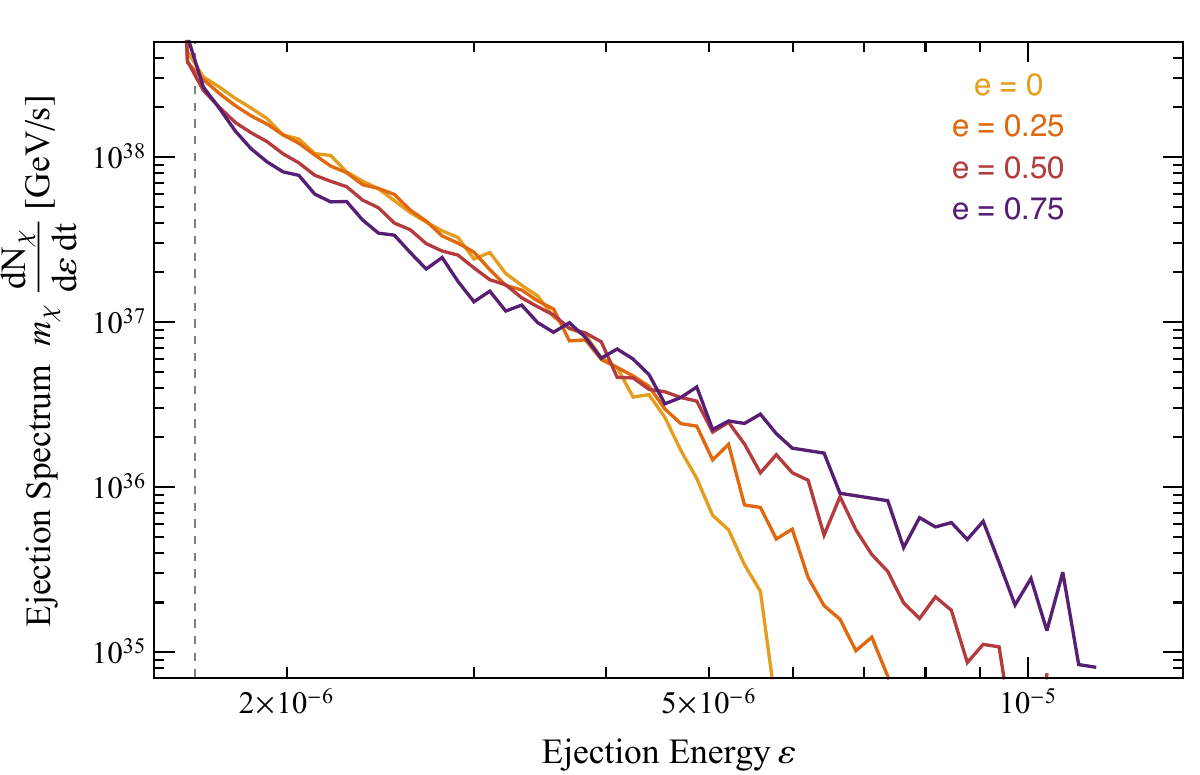}
    \caption{Simulated ejection spectra from equal-mass black hole binaries with total mass $M_{1} + M_{2} = 28 \ M_\odot$, for a fixed period $P = 10^{-2} \ \rm yr$ and various eccentricities. These assume a background DM density of $\rho_\chi = 0.42 \ \rm GeV \, cm^{-3}$, velocity dispersion $\sigma_\chi = 240 \ \rm km/s$ and Galactic escape velocity cutoff $v_{\rm Gal} = 546 \ \rm km/s$ (indicated by the vertical dashed line).}
    \label{fig:eccec_variation}
\end{figure*}

Figure~\ref{fig:eccec_variation} shows the variation of the ejection spectrum with eccentricity, for a fiducial double black hole binary of $10^{-2} \ \rm yr$ period and masses $M_1 = M_2 = 14 \ M_\odot$. Compared to the circular case, eccentric binaries have a suppressed ejection spectrum at lower energies while simultaneously achieving larger ejection energies. This variation can be qualitatively understood in terms of the deflection argument in Sec.~\ref{sec:grav_ejection_analytic}. For a finite eccentricity, both binary components now orbit around each other in elliptical orbits sharing one of the focal points, with the apocenter and pericenter distances given by $a (1 + e)$ and $a (1 - e)$. During most of the orbital period, the binary has a wider orbital extent compared to the circular case, and consequently both objects have slower orbital velocities. As a result, the ejection energy scale is reduced, and achieving a given ejection energy requires closer encounters which are correspondingly less probable. By contrast, during the brief pericenter passage, both components approach closely at high orbital velocities. In this regime, even relatively weak deflections can yield energy gains comparable to, or exceeding, those of the circular case. For this reason, the ejection spectrum subsequently overtakes and extends further in energy compared to the fiducial case. Overall, however, it can be seen that varying eccentricity over most of its allowed range does not significantly alter the ejection spectra for these systems beyond a factor $\sim 2$ in ejection energy. 

\subsection{Mass Ratio}
\label{app:mass_ratio}
\begin{figure*}[t!]
    \centering
    \includegraphics[width=0.9\textwidth]{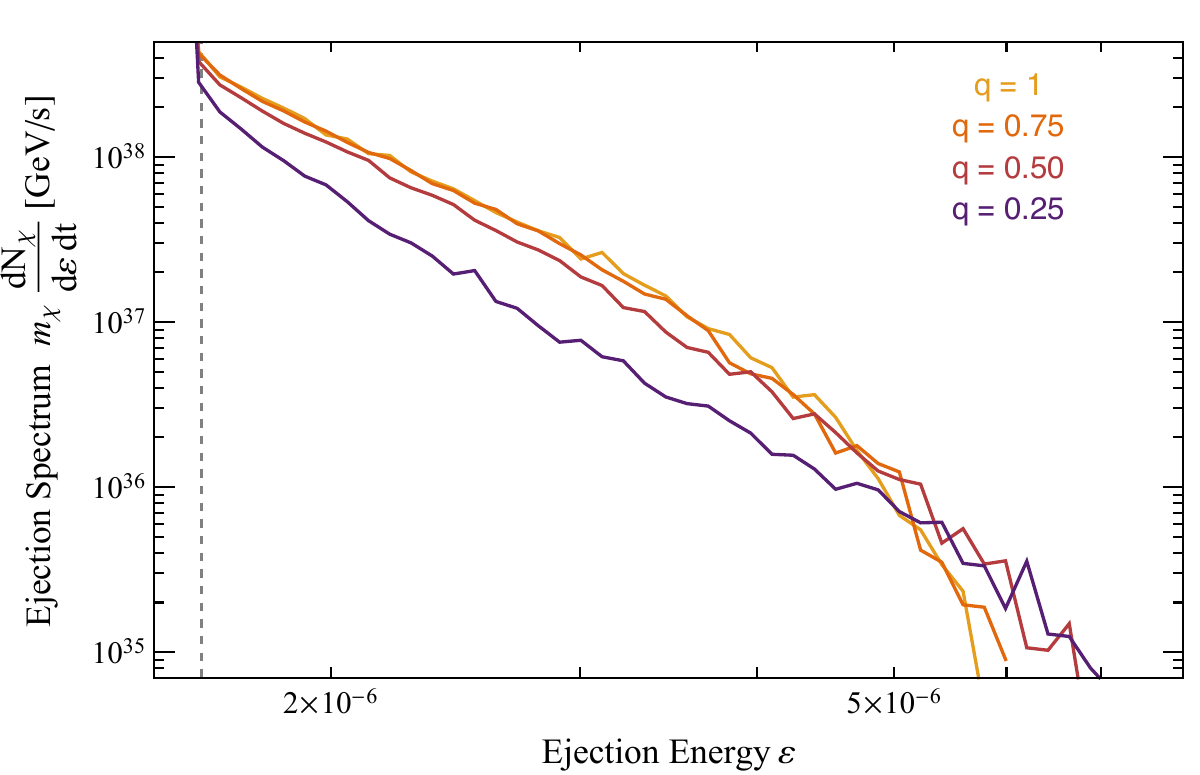}
    \caption{Simulated ejection spectra from circular black hole binaries with total mass $M_{1} + M_{2} = 28 \, M_\odot$ and various mass ratios $q = M_2/M_1$, along with the same assumptions otherwise as for Fig.~\ref{fig:eccec_variation}.}
    \label{fig:mratio_variation}
\end{figure*}

Figure~\ref{fig:mratio_variation} shows the variation of the ejection spectrum with the mass ratio $q = M_2 / M_1$, again for a fiducial double black hole binary of $10^{-2} \ \rm yr$ period and total mass $M_{1} + M_2 = 28 \, M_\odot$. Compared to the equal-mass scenario, as $q$ decreases from unity, we observe a suppressed ejection spectrum combined with a further reach in maximum energy. Similar to the finite eccentricity analysis, this variation can also be qualitatively understood in terms of gravitational deflections. Varying the mass ratio, at fixed total mass and period, leaves the separation between binary components unchanged. However, as the mass ratio $q$ decreases, the lighter companion moves faster about the binary's barycenter, whereas the heavier one wobbles slowly about the barycenter. This implies that the heavy companion is unable to eject particles at relevant energies, given its suppressed orbital velocity. On the other hand, the lighter companion is able to eject particles at higher energies compared to the equal-mass scenario, and therefore the ejection spectrum extends further in energy. However, the ejection rate is now suppressed since one of the binary components cannot significantly contribute to the ejections, whereas its counterpart, being lighter, has a reduced gravitational cross-section for deflections. The overall spectrum is fairly insensitive to the mass ratio except in the extreme mass ratio regime $q \rightarrow 0$, where the spectrum scales down as $M_2^2$. This is discussed further below in the context of Sgr.~A$^\star$. However, for local and bulge binaries, which are comprised of black holes derived from stellar progenitors, the mass ratio cannot be significantly smaller than $\sim 0.5$, and so we conclude that this parameter has a small impact on the corresponding ejection estimates. 

\subsection{Barycenter Motion}
\label{app:syst_motion}
The remaining parameter to consider is the motion of the binary's barycenter relative to the DM halo. As the simulation is performed in the barycenter rest frame, this is parameterized by a wind vector $\mathbf{V}_{\rm w}$ that shifts the initial velocities of all the DM particles. 

\begin{figure*}[t!]
    \centering
    \hspace*{-2.0cm}
    \includegraphics[width=1.25\textwidth]{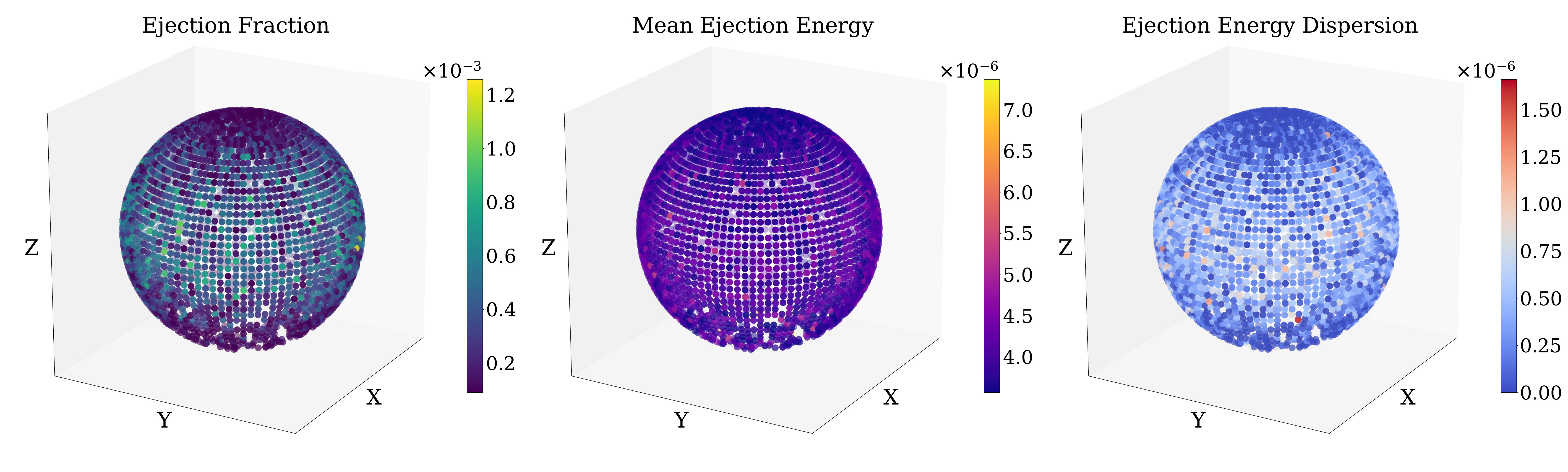}
    \hspace*{-2.0cm}
    \includegraphics[width=1.25\textwidth]{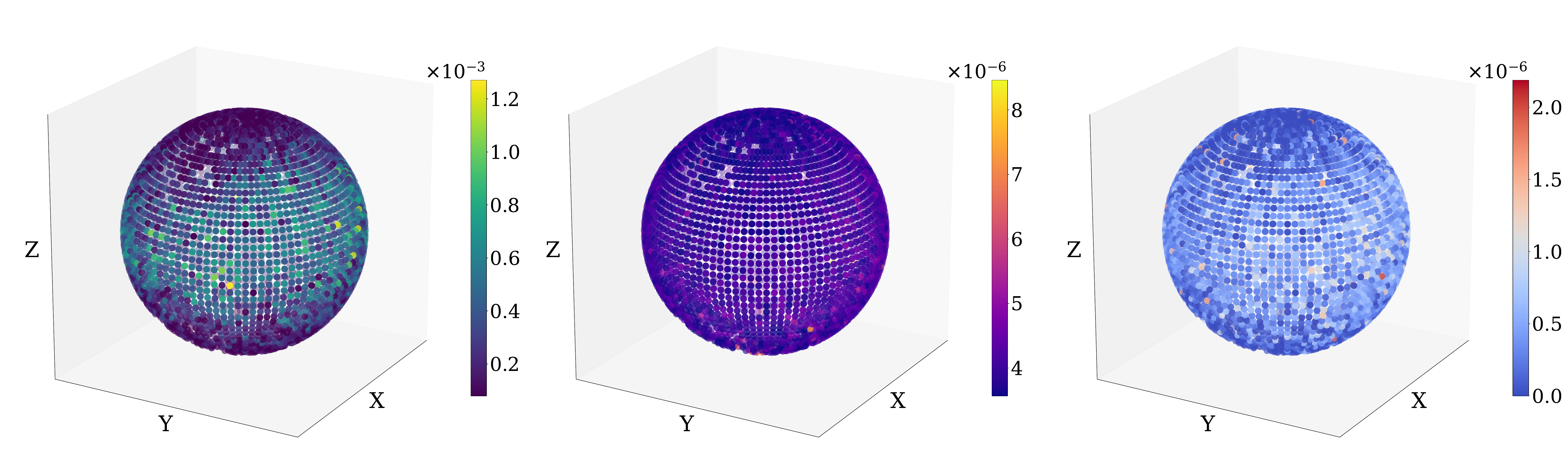}
    \hspace*{-2.0cm}
    \includegraphics[width=1.25\textwidth]{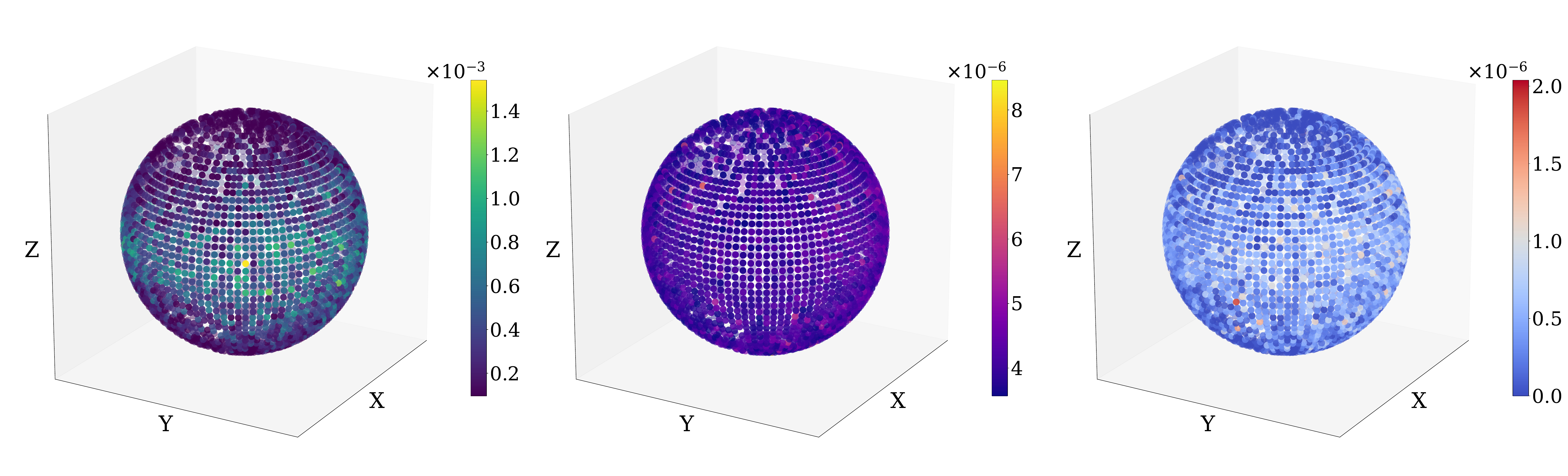}
    \caption{The angular distribution of ejecta as Fig.~\ref{fig:ejecta_ang_dist_no_Vcm}, but now assuming a relative velocity between the binary's barycenter and DM halo of $240 \ \rm km/s$ of magnitude, with angles $90^\circ$ (\textit{top}), $45^\circ$ (\textit{middle}) and $0^\circ$ (\textit{bottom}) relative to the binary plane (the equator in all cases). Empty angular bins imply no ejection was recorded in the full ensemble. In each case, the relative velocity is in the X-Z plane, with the X component pointing from back to front and the Z component from bottom to top.}
    \label{fig:ejecta_ang_dist_with_Vcm}
\end{figure*}

Figure~\ref{fig:ejecta_ang_dist_with_Vcm} shows the angular distribution of the ejecta fraction, average energy, and energy dispersion, for $\mathbf{V}_{\rm w} = 240 \ {\rm km/s} \times  \hat{Z} , \, (\hat{X} + \hat{Z}) / \sqrt{2}\,$, and $\hat{X}$ from top to bottom. As for Fig.~\ref{fig:ejecta_ang_dist_no_Vcm}, the sample has been limited to those particles with $\Delta \varepsilon > 0$ and with final energy above the Galactic escape velocity. The wind magnitude is chosen based on the most likely value predicted by the synthetic population model of Ref.~\cite{Olejak:2019pln}. The orientations are chosen respectively so that, viewed from the halo rest frame, the binary's barycenter velocity is either parallel, at an angle of $45^\circ$, or perpendicular to the orbital angular momentum vector. Compared to the binary being at rest in the halo frame ($cf.$~Fig.~\ref{fig:ejecta_ang_dist_no_Vcm}), ejections are now biased in a direction that correlates with $\mathbf{V}_{\rm w}$. 

When $\mathbf{V}_{\rm w} \, // \ \hat{Z}$, a larger fraction of ejecta are recorded toward $+Z$. This reflects the fact that most encounters are relatively weak: since incoming particles are preferentially injected with velocities pointing in this direction, the resulting ejecta partially retain this initial anisotropy. For the same reason, there is a corresponding deficit in the $-Z$ direction. By contrast, when $\mathbf{V}_{\rm w} \, //  \ \hat{X}$, this anisotropy is drastically shifted. In this case, ejections exhibit a preference within the binary plane in the $+X$ $-$ $+Y$ quadrant, following an approximate vector product rule between binary spin and wind direction. Because the wind lies partially in the orbital plane, ejections now preferentially occur when either companion is facing toward $-X$, which guarantees encounters on their leading side of motion. As discussed in Sec.~\ref{sec:grav_ejection_analytic}, for these interactions, energy gain is guaranteed for any scattering angle. Consequently, ejections predominantly occur towards the $+X$ $-$ $+Y$ quadrant due to the companion, at the time of the encounters, at least partially moving towards $-X$. Compared to the previous case, there is now a dearth of ejected particles in the $\pm Z$ directions, with this gap particularly tilted towards the $+X$ direction. The case $\mathbf{V}_{\rm w} \, // \ (\hat{Z} + \hat{X})$ represents an intermediate regime between these two limiting cases.

As in the case of no barycenter motion ($i.e.$ a binary at rest relative to the halo), we do not observe a strong directional bias for the average energy or dispersion. Upon examining the corresponding spectra as computed using Eq.~\eqref{eq:diff_spec_gen}, we find that the maximum ejection energy is marginally increased due to the finite magnitude of $\mathbf{V}_{\rm w}$, which boosts the initial energy of the ejected particles, but otherwise does not exhibit significant changes relative to the results in the main text beyond a factor of order unity. 

Performing a full scan over these additional three parameters ($i.e.$ orientation and magnitude) makes any estimation of the flux considerably more cumbersome. Since we do not observe a strong variation relative to the assumption that the binary is at rest in the halo frame, we have opted in this work to neglect the binary's motion relative to the halo. As before, we additionally expect such an approximation to average out upon integrating over a full binary population. 

\section{Scaling for Extreme Mass Ratios}
\label{app:M_squared_scaling}
\begin{figure*}[t!]
    \centering
    \includegraphics[width=0.9\textwidth]{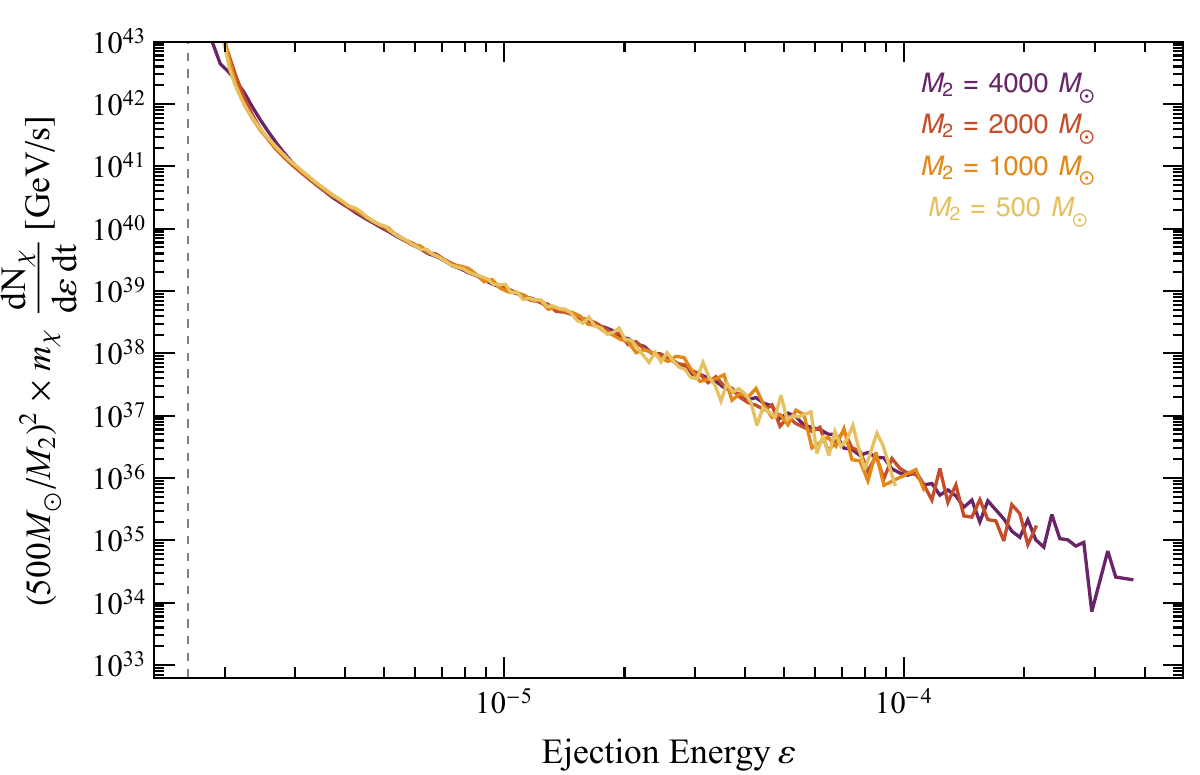}
    \caption{Simulated ejection spectra from Sgr.~A$^\star$ ($M_1 \simeq 4.3 \times 10^6 \ M_\odot$) plus a black hole on a circular, $P = 1 \ \rm yr$ orbit, for various secondary masses as specified. In each case, the contour has been rescaled by a factor $(500 \, M_\odot / M_2)^2$ to illustrate the scaling law with secondary mass in this regime.}
    \label{fig:M2_scaling}
\end{figure*}

Figure~\ref{fig:M2_scaling} shows numerical ejection spectra for a series of systems comprised of Sgr.~A$^\star$ plus a black hole of various masses on a $1$-yr circular orbit, always in the limit $M_2 \ll M_1$. To confirm the $M^2_2$-scaling used in the main text, each curve has been scaled down by a factor $(500 \, M_\odot / M_2)^2$, where $500 \, M_\odot$ is the lowest companion mass we have simulated with our current setup. 

The highest ejection energies we numerically observe correspond to ejection speeds of order $\sim 8300 \ \rm km/s$ for $M_2 = 4000 \, M_\odot$. Since this run effectively corresponds to simulating a restricted 3-body problem, we can directly compare this value to the theoretical maximum set by Eq.~\eqref{eq:delta_epsilon}. The orbital velocity of the companion is $v_{M_2} \simeq 4850 \ \rm km/s$, whereas the initial DM speed upon interacting with the secondary to an excellent approximation is given by Sgr.~A$^\star$'s escape velocity, roughly $v_i \simeq 6860 \ \rm km/s$ at the distance set by the orbital separation. The resulting maximum ejection speed derived for these parameters is $\sim 11,500 \ \rm km/s$. The simulation does not attain the theoretical maximum because it would require a strong deflection produced by an extremely fine-tuned set of initial conditions with vanishingly small probabilities.

We emphasize that the maximum ejection energy is independent of the companion mass in this extreme mass ratio regime, since the orbital velocity of the secondary is effectively determined by the mass of Sgr.~A$^\star$ for a specified period. The absence of observed ejections at higher energies for lighter companions is driven by their intrinsically lower probability of producing such events given our sample size, rather than a suppression of physical origin.

\newpage

\bibliographystyle{jhep}
\bibliography{biblio}

\end{document}